%% file: RFIM_LD.tex
\documentclass[a4paper,11pt]{article}
\usepackage[utf8]{inputenc}
\usepackage{graphicx}
\usepackage{subcaption}
\usepackage[affil-it]{authblk}
\usepackage[a4paper,margin=2.54cm]{geometry}
\input{./header}

\newcommand{\T}{\theta}
\newcommand{\hm}{\hat{m}}

\newcommand{\vm}{\v{m}}

\newcommand{\vhm}{\hat{\v{m}}}
\newcommand{\ihm}{i\hat{m}}
\newcommand{\sv}{\gv{\s}}
\newcommand{\eqr}[1]{eq.~(\ref{#1})}

\title{Large Deviations of the Finite-Time Magnetization of the Curie-Weiss Random Field Ising Model}

\author{
Pierre Paga\thanks{pierre.paga$@$ kcl.ac.uk}
and
Reimer K\"{u}hn\thanks{Reimer.kuehn$@$ kcl.ac.uk}
}
\affil{Department of Mathematics,\\King's College London ,UK\\}

\begin{document}

\maketitle

\abstract{
We study the large deviations of the magnetization at some finite time in the Curie-Weiss Random Field Ising Model with parallel updating.
While relaxation dynamics in an infinite time horizon gives rise to unique dynamical trajectories (specified by initial conditions and governed by first-order dynamics of the form $m_{t+1}=f(m_t)$), we observe that the introduction of a finite time horizon and the specification of terminal conditions can generate a host of metastable solutions obeying \textit{second-order} dynamics. We show that these solutions are governed by a Newtonian-like dynamics in discrete time which permits solutions in terms of both the first order relaxation (``forward'') dynamics and the backward dynamics $m_{t+1} = f^{-1}(m_t)$. Our approach allows us to classify trajectories for a given final magnetization as stable or metastable according to the value of the rate function associated with them. We find that in analogy to the Freidlin-Wentzell description of the stochastic dynamics of escape from metastable states, the dominant trajectories may switch between the two types (forward and backward) of first-order dynamics.
}

\section{Introduction}

Disordered systems are characterized by large fluctuations in their physical observables induced by the presence of quenched randomness. 
This creates a situation  where not only average physical quantities and the variances of fluctuations about these averages are of interest, but their entire distribution has physical relevance. 
In this context the study of large deviations arises naturally \cite{touchette2009large, Biroli07}, and their applicability is very general \cite{LebowitzSpohn}. 
As a result, the topic has attracted much attention in recent years in a diversity of fields including spin glasses \cite{morone2014large, parisi2007large,Gal09}, kinetically constrained models \cite{garrahan2007dynamical,bodineau2012activity}, random matrix theory \cite{BGKP,VMB, KPC}, and epidemic spreading on networks \cite{Greven94, ABDZ}.

The study of large deviations also has economic importance \cite{Kluppelberg}: it is vital for example for an insurer to estimate the likelihood of situations in which large numbers of indemnities may have to be paid at once. 
From the regulator's perspective, understanding how often large-scale crises may occur in a given regulatory scenario is key to proper risk-management policy. 
But while attempts at studying rare events in semi-realistic credit risk models have been made \cite{PK}, computing large-deviation functions has proven too difficult due in part to the presence of quenched disorder. 

This difficulty motivates the study of simple models with quenched disorder for which a complete analysis of dynamical large deviation properties is possible and which could therefore be used as testbeds for approximation methods. Of such simple models, the Ising model is the prototypical example.
Recently, the large deviations of the Ising model have received great attention, e.g. studies of the large deviations of the energy in the 1-d Ising chain, or of the activity (the number of times a spin flips in a given trajectory), as in \cite{RS10,loscar2011thermodynamics}.
A natural extension of the Ising model to include quenched disorder is the random-field Ising model (RFIM). The large deviations of the equilibrium magnetization have long been well understood \cite{dMP91,lowe2013large} and have proven a useful tool in understanding the phase diagram of the model. To the best of our knowledge however no studies of the large deviations of the finite-time magnetization in the RFIM have been made even in its simplest (Curie-Weiss) description. Yet it is of clear interest to have a simple but non-trivial model with disorder that admits closed-form solutions against which to test our intuition and approximate methods.

In this article we thus treat the large deviations of the finite-time magnetization in the Curie-Weiss Random Field Ising Model with parallel updating. Using generating functional methods, we derive equations of motion which include the value of the magnetization at a finite time as a constraint and find that it can be recast in a language reminiscent of Newtonian dynamics in discrete time. 
This language also arise naturally as a zero-noise limit of Langevin dynamics where the magnitude of the noise scales like $N^{-1/2}$, with $N$ the system size. 

The remainder of this paper is organised as follows: we introduce the model and our main notations, then introduce a discrete time path integral formalism from which we derive saddlepoint equations. We analyse these equations and find non-trivial fixed points (i.e. fixed points which are not equilibrium solutions) and find them to be elliptical. We then derive Newtonian-like dynamics for the system at the saddlepoint. We derive rate functions and compare them with explicit simulations for small numbers of time steps ($T=50$) in the ferromagnetic parts of the phase diagram. 

\section{Model}
We investigate the parallel dynamics of the Curie-Weiss RFIM, where the transition probabilities between configuration $\cup{\s_i}_{i=1,N}$ and $\cup{\s'_i}_{i=1,N}$ is given by
\begin{align}
\label{eq:transition}
 W\rp{\cup{\s'_i}|\cup{\s_i}} = \prodl{i}{}\f{e^{\b\s'_i\rp{Jm + h\T_i}}}{2\cosh \sp{\b\rp{Jm + h\T_i}}}
\end{align}
where $m = N^{-1}\suml{i}{}\s_i$ is the magnetization of configuration $\cup{\s_i}$, the $\T_i$ are random fields which take value in $\pm 1$ with probability $(p_\T,1-p_\T)$ and $h$ represent the strength of these random fields. 
We take $J=1$ without loss of generality, and assume the initial spins are i.i.d: $P(\cup{\s_{i0}}) = \prodl{i}{}p_0\rp{\s_{i0}}$. We parametrize the distribution of initial spins as $p_0(\s) = \tfrac{1+ \s r_0}{2}$. 
In what follows we will write $\avg{\cdots}$ the average with respect to the dynamics in \eqr{eq:transition}, and by $\avg{\cdots}_{\T}$ the average with respect to the random field distribution.

With such a setup, the probability of a sequence of configurations $\cup{\s_{it}}$ is given by
\begin{align}
\label{eq:proba:seq}
 P\rp{\cup{\s_{it}}} = \prodl{t=1}{T}W\rp{\cup{\s_{it}}|\cup{\s_{i(t-1)}}} \prodl{i}{}p_0\rp{\s_{i0}} \ .
\end{align}
At large times, the probability distribution of the magnetization is well-known as the equilibrium dynamics are governed by Peretto's pseudo-Hamiltonian  (\cite{Peretto1984,SkantzosCoolen00})
\begin{align}
 H_{\b}(\v{\s}) = -\f{1}{\b}\suml{i}{}\log\sp{2\cosh\rp{\b\sp{m+h\T_i}}} - h\suml{i}{}\s_i\T_i \ ,
\end{align}
which gives the rate function
\begin{align}
\label{eq:Peretto}
 I_{\b}(m) =& \lim_{N\ra\infty}-\f{1}{N}\log P\rp{\f{1}{N}\suml{i}{}\s_i = m}
 \\=&
 \sup_{x}\cup{xm - \avg{\log\sp{\cosh\rp{\b h\T + x}}}_{\T}} - \avg{\log\sp{\cosh\rp{\b \sp{h\T + m}}}}_{\T} + I_0 \ .
\end{align}
Here $I_0$ is a constant chosen such that $\uos{m}{}{\min}\cup{I_{\b}(m)} = 0$.
\begin{figure}
 \begin{subfigure}{0.5\textwidth}
 \includegraphics[width=\textwidth]{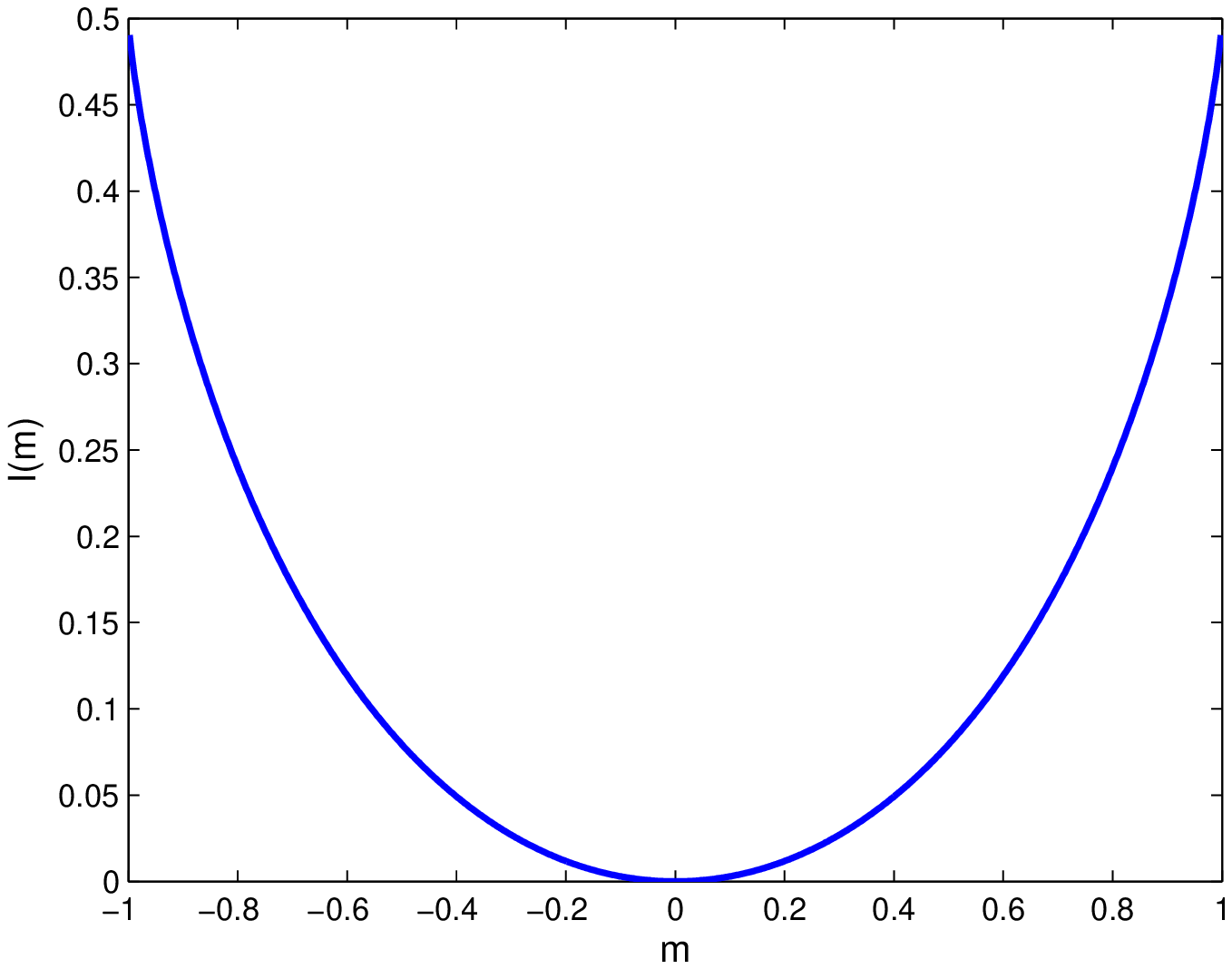}
 \caption{}
   \label{fig:Peretto.1}
 \end{subfigure}
 \begin{subfigure}{0.5\textwidth}
 \includegraphics[width=\textwidth]{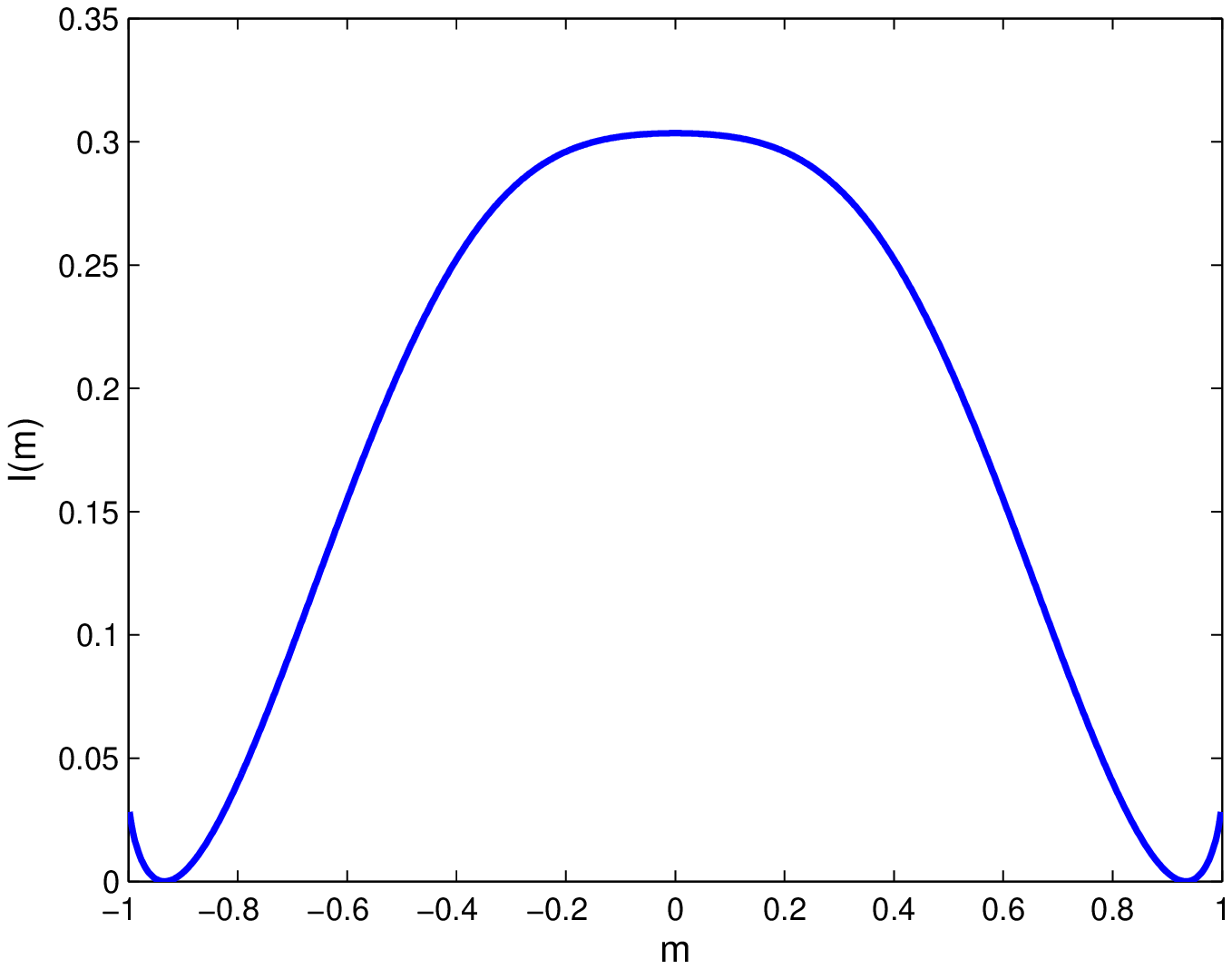}
 \caption{}
  \label{fig:Peretto.2}
 \end{subfigure}
 \caption{
 Peretto rate functions in the paramagnetic ((a), $\b = 0.67$, $h=0.2$)  and ferromagnetic ((b), $\b = 2.5$, $h=0.4$) parts of the phase diagram
 }
 \label{fig:Peretto}
\end{figure}
But the understanding of the equilibrium distribution alone may not be sufficient in many problems, and provides no way to estimate e.g. the equilibration time. To remedy this, we compute the large deviation functions of the magnetization at finite times.

\section{Dynamics of the magnetization}
We seek to compute the large deviation rate function of the magnetization $m_T$ for some finite $T$.
\subsection{Trajectory probability}
Writing $\v{m} = (m_t)_{t=0, \cdots, T}$, magnetization path probabilities are given by
\begin{align}
 P(\v{m}) 
 =& 
 \avg{\prodl{t}{}\delta\rp{m_t - \frac{1}{N}\suml{i}{}\s_{it}}}
\\=&
 \suml{\cup{\s_{it}}}{}\prodl{i}{}p_0\rp{\s_{i0}}\prodl{t=1}{T}\sp{\f{e^{\b \rp{\s_{it}m_{t-1} + h\T_i\s_{it}}}}{2\cosh\sp{\b\rp{m_{t-1} + h\T_i}}}}
 \prodl{t=0}{T}\delta\rp{m_t - \frac{1}{N}\suml{i}{}\s_{it}}
 \intertext{We insert the integral representation $\delta(x) = \int \tfrac{\d{y}}{2\pi}e^{ixy}$ at every time step:}
 P(\v{m}) =&
 \suml{\cup{\s_{it}}}{}e^{N\b \suml{t=1}{T}m_tm_{t-1}}
\prodl{i}{}p_0\rp{\s_{i0}}\prodl{t=1}{T}
 \f{e^{\b h\T_i\s_{it}}}{2\cosh\sp{\b\rp{m_{t-1} + h \T_i}}}
 \notag\\&
 \times\sp{\f{N}{2\pi}}^{T+1} \int \d{\v{\hm}} 
 \exp\cup{
 -i\v{\hm}\cdot\rp{N\v{m} - \suml{i}{}\gv{\s}_i}
 }
\intertext{where we write $\vhm = (\hm_t)_{t=0, \cdots, T}$ and $\gv(\s)_i = (\s_{it})_{t=0, \cdots, T}$. 
This representation allows for the decoupling of the sum over microstates $\cup{\s_{it}}$ w.r.t the sites i, and thus $P(\v{m})$ can be expressed as}
P(\v{m})=&
\sp{\f{N}{2\pi}}^{T+1}\int \d{\v{\hm}} 
\exp\cup{ - iN\v{\hm}\cdot\v{m} + N\b \suml{t=1}{T} m_{t}m_{t-1}
-\suml{i}{}\suml{t=1}{T}\log \sp{2\cosh\rp{\b\sp{m_{t-1} + h \T_i}}}
}
\notag\\
&\qquad
\prodl{i}{}
\suml{\cup{\s_{it}}}{}
p_0(\s_{i0})e^{i\hm_0}\exp\cup{\s_{t}\rp{\T_i + i\hat{m}_{t}}}
\intertext{and can be expressed as a discrete time path integral}
P(\v{m})=&
\sp{\f{N}{2\pi}}^{T+1}\int \d{\v{\hm}} \exp\cup{- iN\v{\hm}\cdot\v{m} + \b N\suml{t=1}{T}m_tm_{t-1}
+ N\suml{t=0}{T}\avg{\log\sp{Z_t(\T)}}_{\T} } 
\ ,
\label{eq.path.integral}
\end{align}
with
\begin{align}
 Z_0(\T) =& \sqrt{1-r_0^2}\f{\cosh\sp{\rho + \ihm_0}}{\cosh\sp{\b\rp{m_0+h\T}}}\ ,
\\
 Z_t(\T) =& \f{\cosh\sp{i\hat{m}_t + \b h\T}}{\cosh\sp{\b\rp{m_t + h\T}}} \ , \quad 1\leq t \leq T-1 
\\
 Z_T(\T) =& \cosh\sp{i\hat{m}_T + \b h\T}
\end{align}
where $\rho = \tanh^{-1}\rp{r_0}$.

\subsection{Final magnetization}
To find the marginal probability of the final magnetization $m_T$, we integrate over the rest of the trajectory. This gives
\begin{align}
 P\rp{m_T}
 \propto
 \int \sp{\prodl{t=0}{T-1}\d{m_t}\d{\hm_t}}\d{\hat{m}_T} \exp\cup{-N\Omega\rp{\v{m},\v{\hm}}}
\label{eq:final.magnetization.probability}
 \end{align}
 with
 \begin{align}
 \label{eq:Omega.def}
 \Omega\rp{\v{m},\v{\hm}} = &
\suml{t=0}{T}i\hm_tm_t - \b\suml{t=1}{T}m_{t-1}m_t
 -\suml{t=0}{T}\avg{\log\sp{Z_t(\T)}}_{\T} \ .
\end{align}
The integral in \eqr{eq:final.magnetization.probability} can at large $N$ be evaluated by the saddlepoint method. 
At the saddlepoint we have the following conditions:
\begin{align}
 \intertext{at $t=0$}
 m_0 =& \tanh\sp{\rho + i\hat{m}_0}
 \label{eq.motion.01}
 \\
 i\hat{m}_0 =& \b\rp{m_{1} - \avg{\tanh\sp{\b\rp{m_0+\T}}}_{\T}}
  \label{eq.motion.02}
 \\
\intertext{for $1\leq t\leq T-1$:}
  m_t =& \avg{\tanh\sp{i\hat{m}_t + \b h \T}}_{\T}
 \label{eq.motion.03}
 \\
 i\hat{m}_t =& \b\rp{m_{t-1} + m_{t+1} - \avg{\tanh\sp{\b\rp{m_t+\T}}}_{\T}}
  \label{eq.motion.04}
 \\
  \intertext{and at $t=T$}
 m_T =& \avg{\tanh\sp{i\hat{m}_T + \b h \T}}_{\T}
 \label{eq.motion.05}
\end{align}
Thus the rate function for the final magnetization $m_T$ reads
\begin{align}
 I(m) = -\lim_{N\ra\infty}\f{1}{N}\log P(m_T=m) = \Omega\rp{\v{m}^*,\vhm^*}|_{m_T^*=m}
\end{align}
where starred quantities denote saddle-point values.

\subsection{Finite-time solutions}
To simplify our expressions, we introduce the notations
\begin{align}
 f(x) =& \avg{\tanh\sp{\b\rp{x+h\T}}}_\T
 \\
 f_0(x) =& \tanh\sp{\rho + \b x} \ ,
 \intertext{likewise}
 F(x) =& \avg{\log\sp{\cosh\rp{\b\rp{x+h\T}}}}_\T
 \\
 F_0(x) =& \f{1}{2}\log\rp{1-r_0^2} + \log\sp{\cosh\rp{\rho + \b x}} \ ,
 \intertext{finally,}
 \t{f}(x) =& f(x) + f^{-1}(x) \ ,
 \\
 \t{f}_0(x) =& f^{-1}_0(x) + f(x) \ ,
\end{align}
Using these notations, the equations of motion (\ref{eq.motion.01}-\ref{eq.motion.05}) are rewritten as
\begin{align}
 m_0 =& f_0\rp{i\b^{-1}\hat{m}_0}
 \label{eq.motion.11}
 \\
 i\hat{m}_0 =& \b\rp{m_{1} - f(m_0)}
  \label{eq.motion.12}
 \\
 m_t =& f\rp{i\b^{-1}\hat{m}_t}
 \label{eq.motion.13}
 \\
 i\hat{m}_t =& \b\rp{m_{t-1} + m_{t+1} - f(m_t)}
  \label{eq.motion.14}
 \\
 m_T =& f\rp{i\b^{-1}\hat{m}_T}
 \label{eq.motion.15}
\end{align}
We can insert the equations for the $i\hat{m}_t$ quantities in the equations for the $m_t$, and using the fact that both $f$ and $f_0$ are invertible (see appendix) we obtain
\begin{align}
 m_1 =& \t{f}_0\rp{m_{0}}
 \label{eq.motion.1}
 \\
 m_{t+1} + m_{t-1} =& \t{f}\rp{m_t}
 \label{eq.motion.2}
 \\
 i\hat{m}_T =& f^{-1}\rp{m_T} \ .
 \label{eq.motion.3}
\end{align}
and we can rewrite the $\Omega$ function as
\begin{align}
\label{eq:Omega}
 \Omega\rp{\vm,\vhm} = &
\suml{t=0}{T}i\hm_tm_t - \b\suml{t=1}{T}m_{t-1}m_t
+F(m_0) - F_0(i\b^{-1}\hm_0)
\notag\\
&+ \suml{t=1}{T-1}\sp{F(m_t) - F(i\b^{-1}\hm_t)} 
- F(i\b^{-1}\hm_T)
\\
=&
\b\suml{t=1}{T}m_t\rp{m_{t-1}-f^{-1}(m_t)}
+\b m_0f_0^{-1}\rp{m_0}
+F(m_0) - F_0(f^{-1}(m_0))
\notag\\
&+ \suml{t=1}{T-1}\sp{F(m_t) - F(f^{-1}(m_t))}
 - F(f^{-1}(m_T)) \ .
\end{align}

Solutions to \eqr{eq.motion.2} can be parametrized via any choice of two points on the trajectory, although not all choices lead to physical trajectories (i.e. $\abs{m_t}<1$ for all $t$). 
With the additional constraint of \eqr{eq.motion.1} and since $m_T$ is fixed, we expect there to be only a finite number of possible solutions. 
Equation (\ref{eq.motion.3}) meanwhile has no effect on the trajectories themselves, it only intervenes in the computation of $\Omega$.

Moreover, we notice that \eqr{eq.motion.2}, which can be considered the proper ``equation of motion'' of the system, include the ``unconstrained'' dynamics as solutions: the unconstrained (or average) equations of motions being
\begin{align}
 m_{t+1} = f(m_t) \ .
\end{align}
Moreover, \eqr{eq.motion.2} is time-reversal invariant, and thus the time-reversed trajectories
\begin{align}
 m_{t+1} = f^{-1}(m_t)
\end{align}
are also valid solutions when far from $t=0$.

\section{Effective dynamics}

\subsection{Quasi-Newtonian dynamics}
\label{Newton}
Given the equations (\ref{eq.motion.1}-\ref{eq.motion.2}), the dynamics of the system can be recast into a form reminiscent of Newtonian dynamics:
\begin{align}
\label{eq:sec-order}
\Delta^2_t m  \equiv \rp{m_{t+1} - m_t} - \rp{m_{t}-m_{t-1}} = k(m_t) \ ,
\end{align}
where we recognize in the left-hand side a discrete time second derivative and write
\begin{align}
k(x) = \t{f}(x) - 2x\ .
\end{align}

\begin{figure}
\centering
\begin{subfigure}{0.45\textwidth}
 \includegraphics[width=\textwidth]{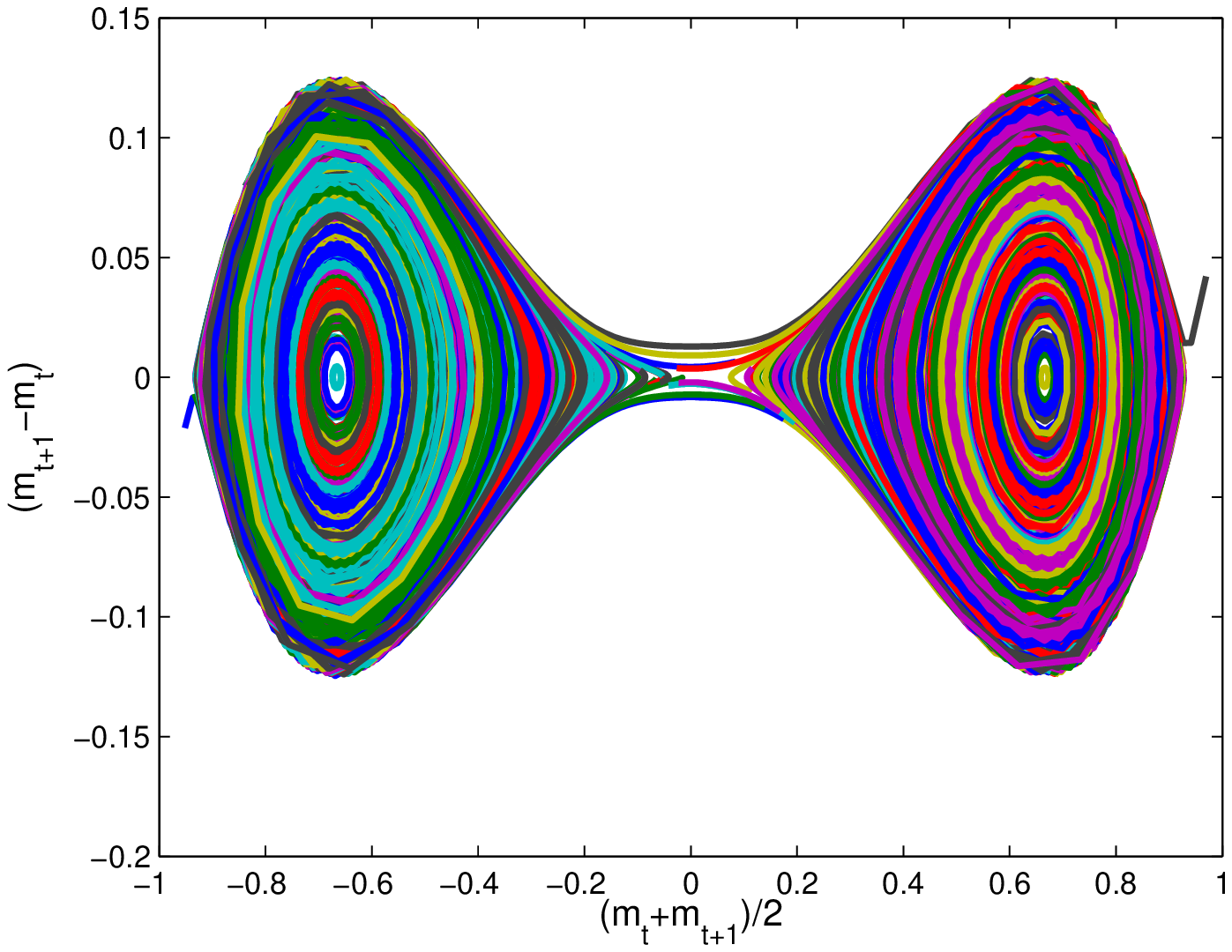}
 \caption{}
\end{subfigure}
\begin{subfigure}{0.45\textwidth}
 \includegraphics[width=\textwidth]{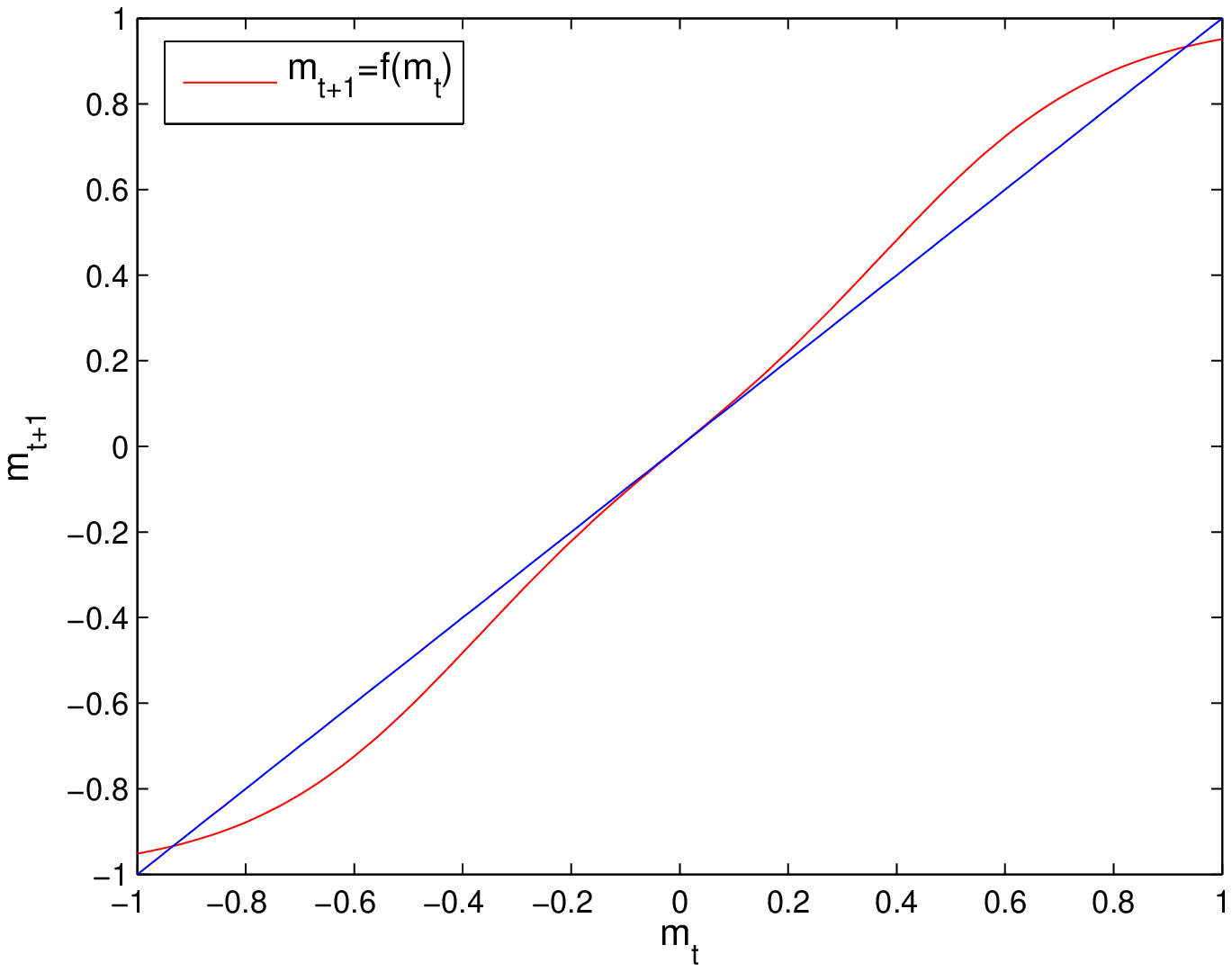}
 \caption{}
\end{subfigure}
\\
\begin{subfigure}{0.45\textwidth}
 \includegraphics[width=\textwidth]{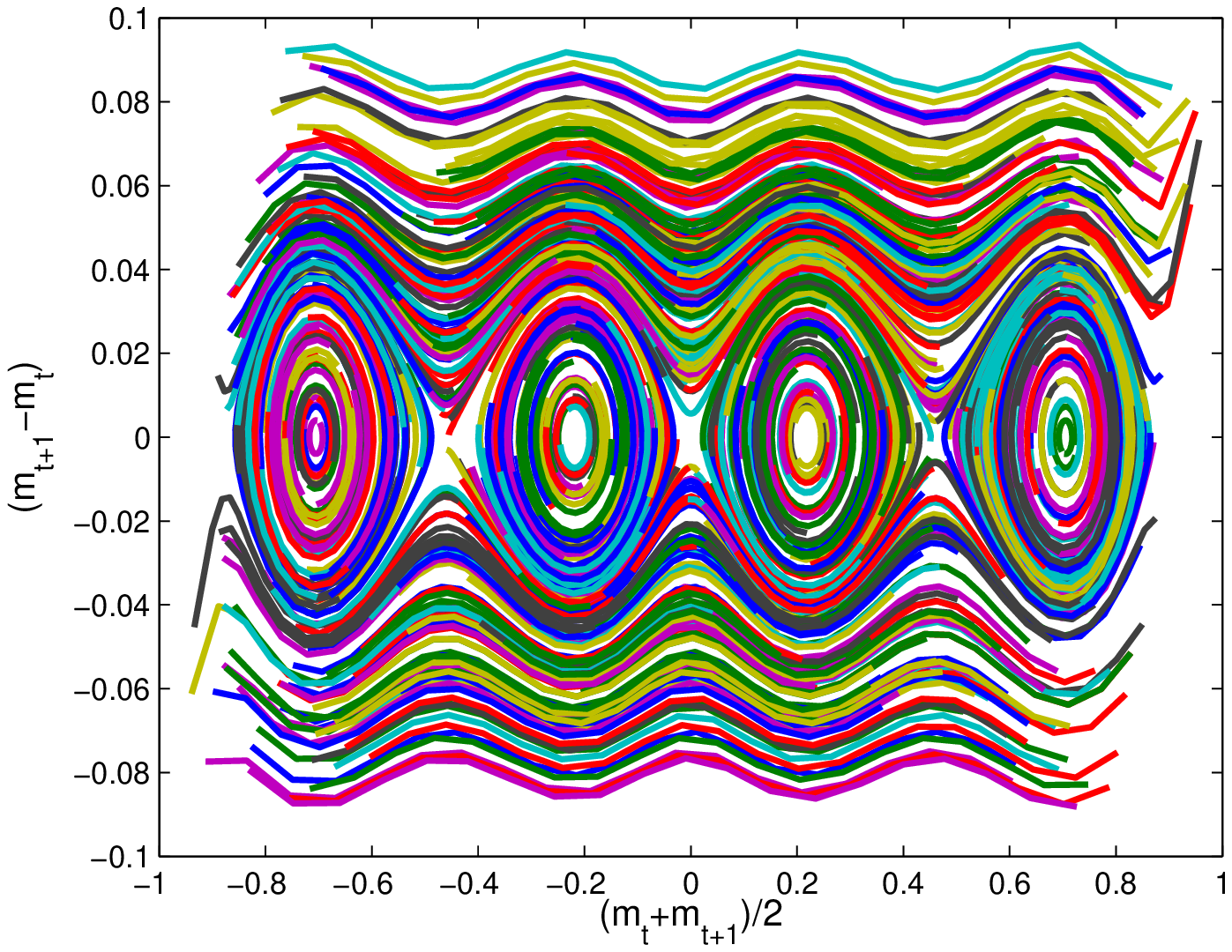}
 \caption{}
\end{subfigure}
\begin{subfigure}{0.45\textwidth}
 \includegraphics[width=\textwidth]{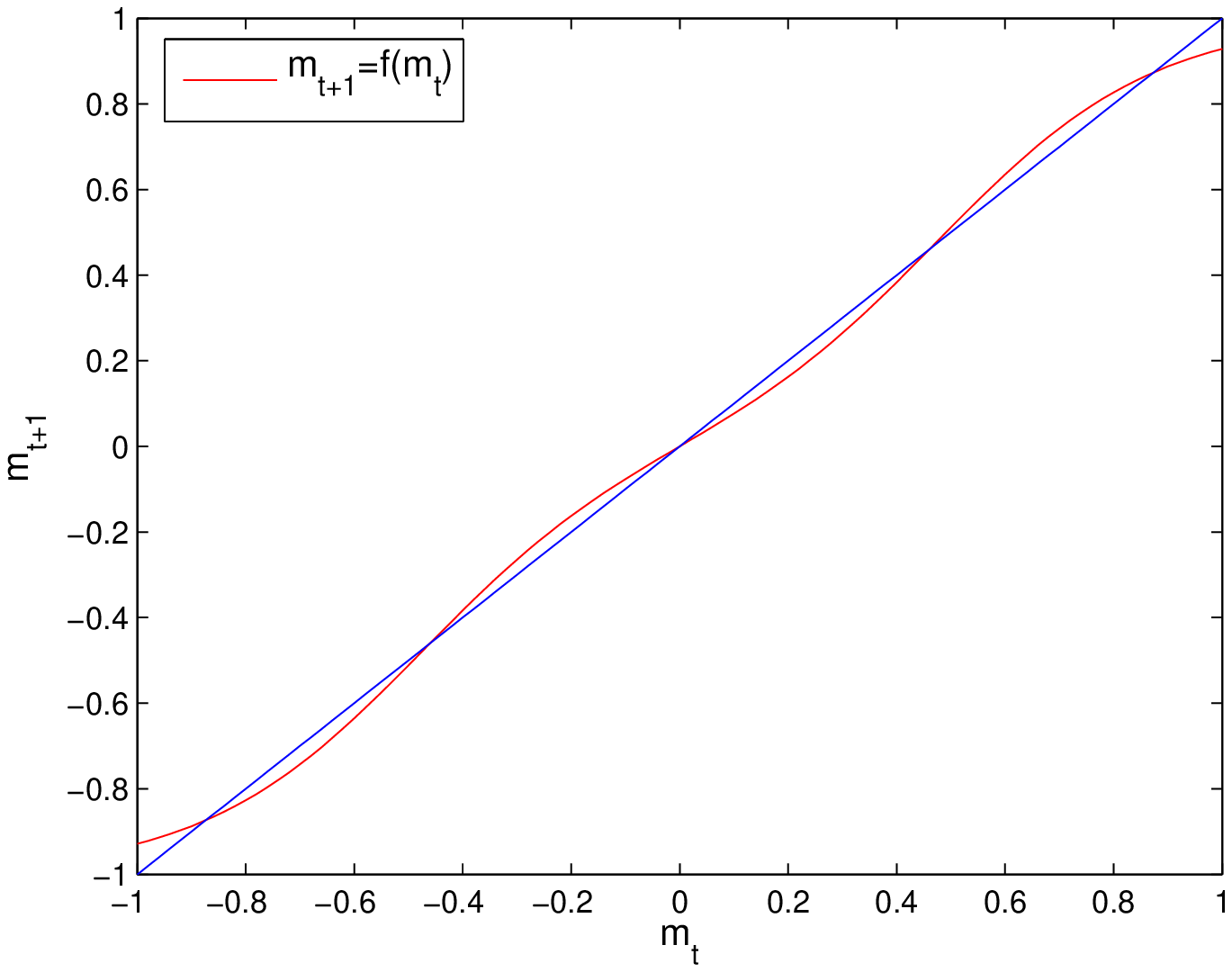}
 \caption{}
\end{subfigure}
\caption{
Phase portraits of trajectories described by \eqr{eq.motion.2} ((a) and (c)) and maps of the relaxation dynamics $m_{t+1}=f(m_t)$ together with the diagonal $m_{t+1}=m_t$ ((b) and (d)) for the parameter settings:
(a-b): $\b=2.5,h=0.4$ (ferromagnetic phase),
(c-d): $\b=2.5,h=0.485$ (ferromagnetic phase with metastable state at $m=0$)
}
\label{fig:phase.portraits.1}
\end{figure}

\begin{figure}
\centering
\begin{subfigure}{0.45\textwidth}
 \includegraphics[width=\textwidth]{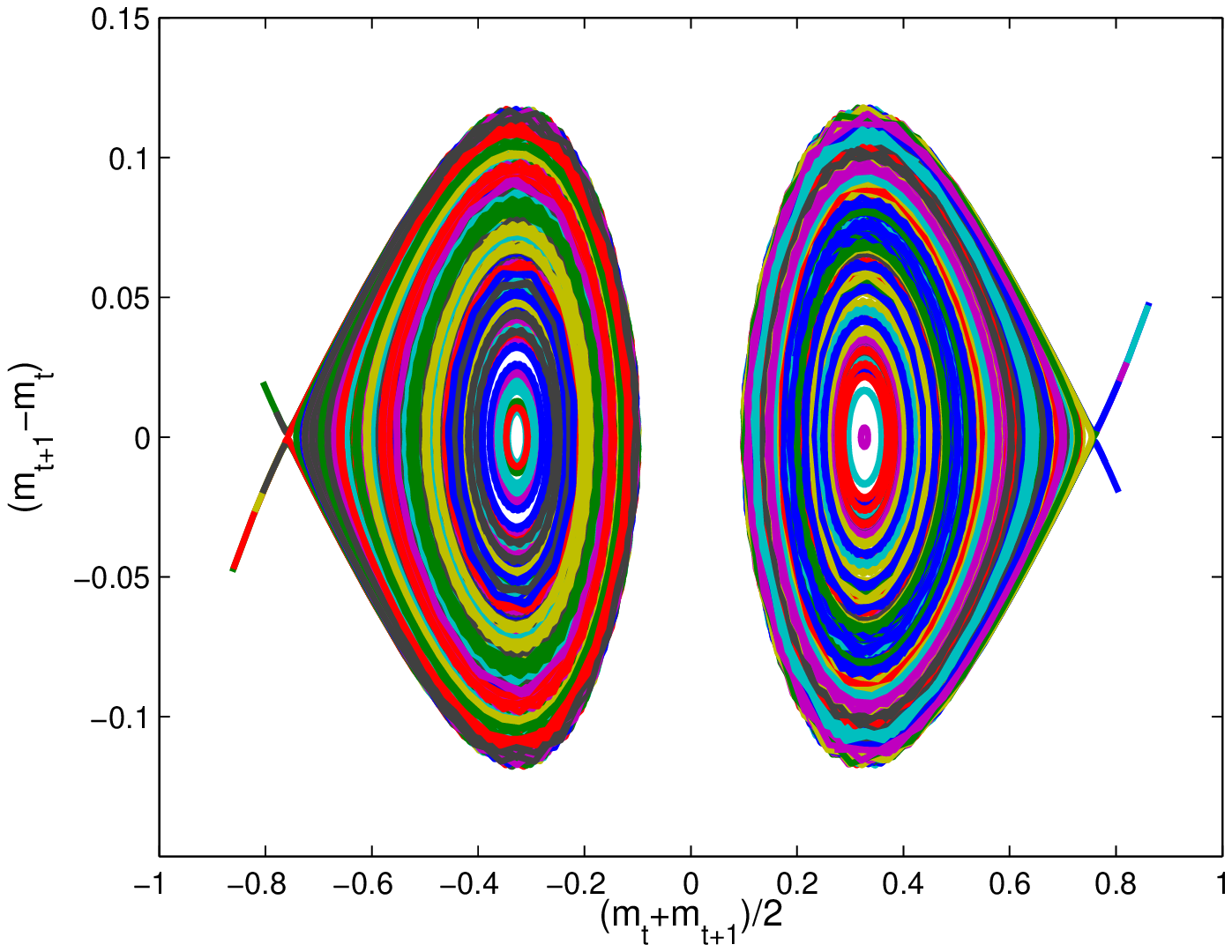}
 \caption{}
\end{subfigure}
\begin{subfigure}{0.45\textwidth}
 \includegraphics[width=\textwidth]{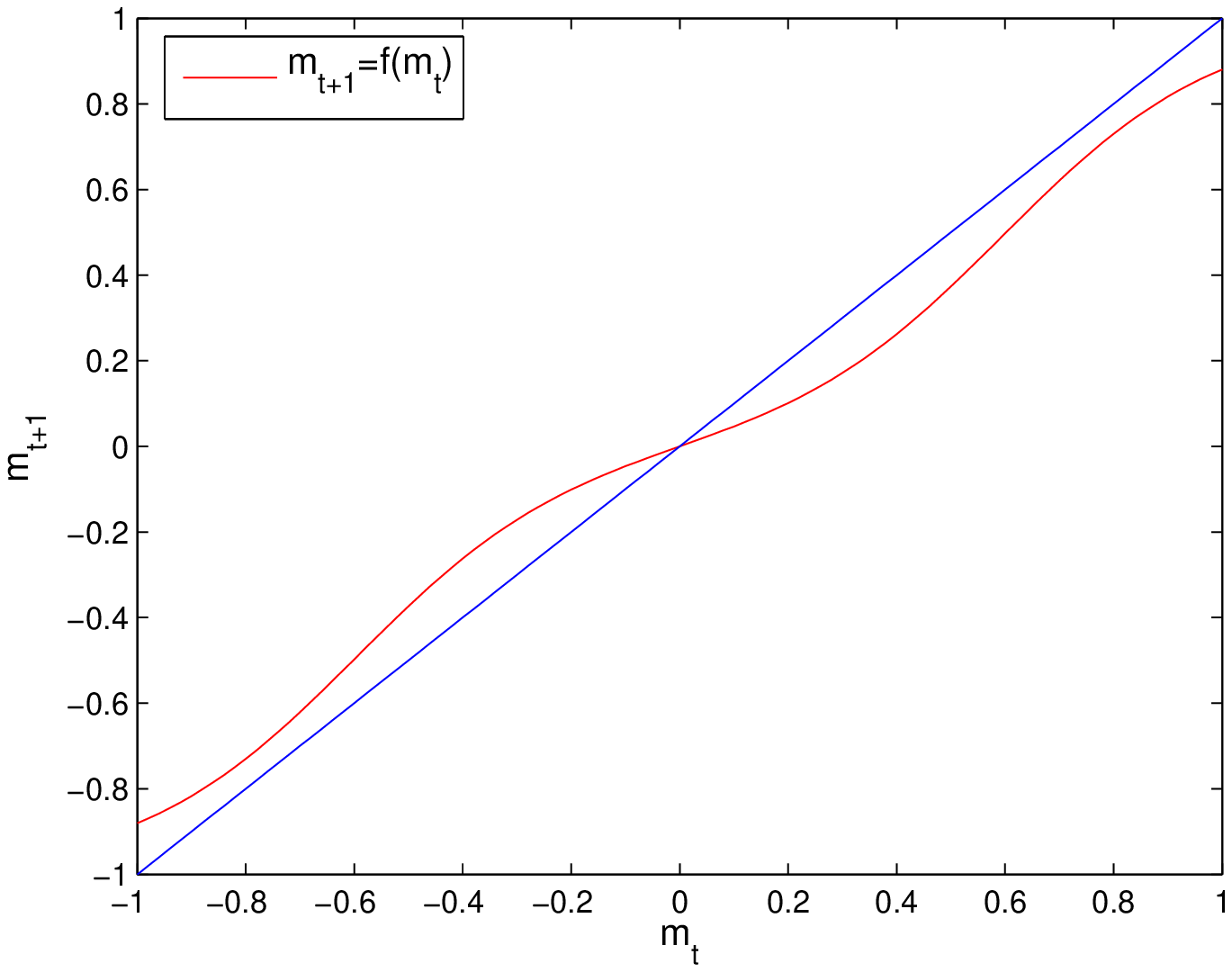}
 \caption{}
\end{subfigure}
\\
\begin{subfigure}{0.45\textwidth}
 \includegraphics[width=\textwidth]{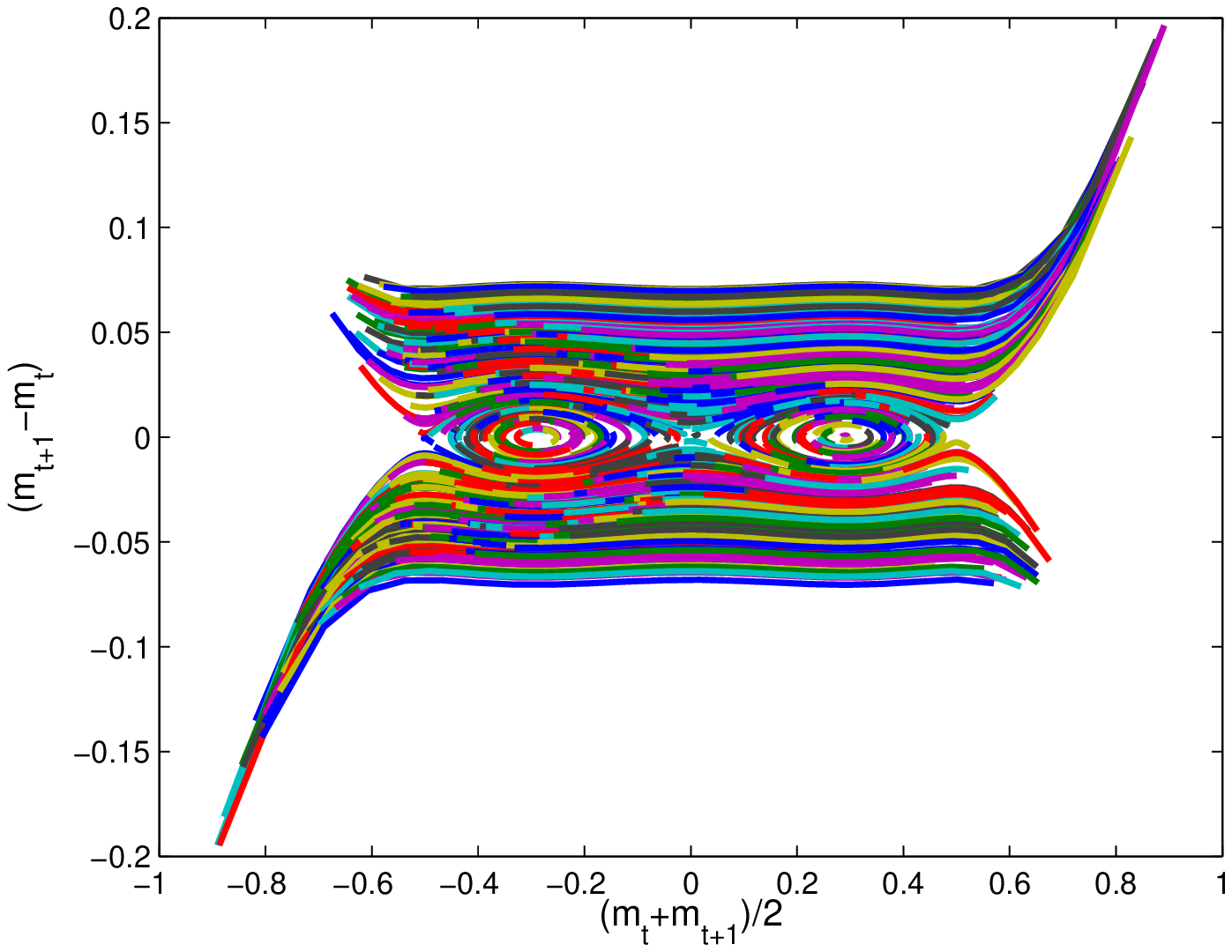}
 \caption{}
\end{subfigure}
\begin{subfigure}{0.45\textwidth}
 \includegraphics[width=\textwidth]{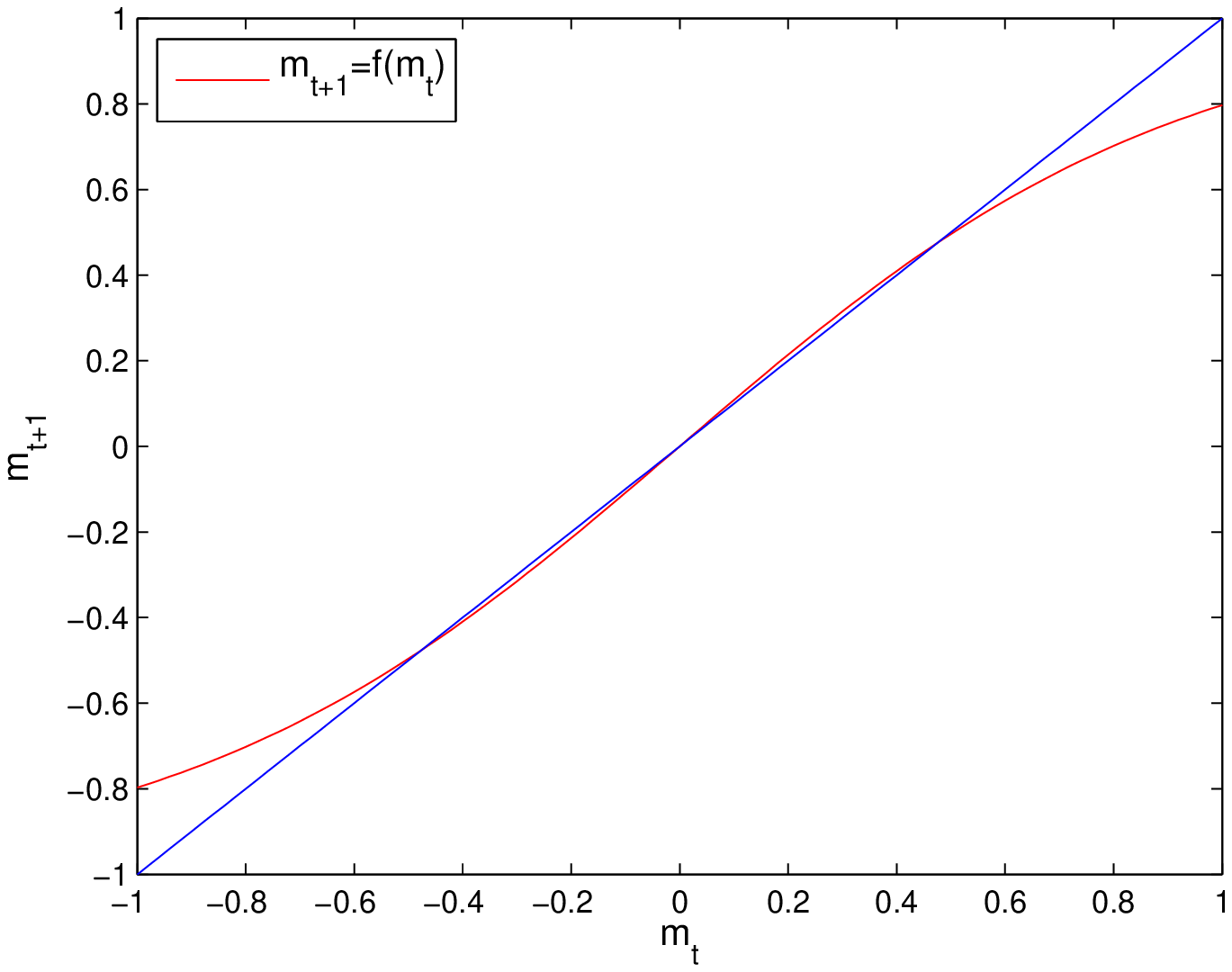}
 \caption{}
\end{subfigure}
\caption{
Phase portraits of trajectories described by \eqr{eq.motion.2} ((a) and (c)) and maps of the relaxation dynamics $m_{t+1}=f(m_t)$ together with the diagonal $m_{t+1}=m_t$ ((b) and (d)) for the parameter settings:
(a-b): $\b=2.5,h=0.6$ (paramagnetic phase)
(c-d): $\b=1.1,h=0.1$ (ferromagnetic phase near a second-order phase transition)
} 
\label{fig:phase.portraits.2}
\end{figure}

In figs.~\ref{fig:phase.portraits.1}-\ref{fig:phase.portraits.2} we plot phase portraits of trajectories corresponding to solutions of \eqr{eq.motion.2}. 
We compute the trajectories by specifying $m_0$ and $m_1$, taken from a uniform grid in $[-1,1]^2$, and computing the values of the subsequent magnetization according to \eqr{eq.motion.2}. 
A subset of these solutions represent solutions of the full system (\ref{eq.motion.1})-(\ref{eq.motion.2}) for suitably chosen $m_T$ and $r_0$.
Additionally, we plot the relaxation dynamics map $m_{t+1} = f(m_t)$ and the diagonal $m_{t+1}=m_t$ to highlight the corresponding equilibrium fixed points.
Since we are dealing with discrete dynamics, it is natural to plot the ``momentum'' $m_{t+1}-m_t$ as a function of the average position $\tfrac{m_{t}+m_{t+1}}{2}$ rather than $m_t$ or $m_{t+1}$ to avoid a tilting of the phase portraits. 
The potential wells appear clearly, with up to four potential wells for $\b=2.5, h=0.485$ in the regime where a ferromagnetic phase coexists with a metastable paramagnetic phase.

\subsection{Energy conservation}

Since eq.~(\ref{Newton}) is reminiscent of Newtonian dynamics, we expect it to follow a discrete form of energy conservation. 
Indeed, multiplying both sides by $(m_{t+1}-m_{t-1})/2$ we have
\begin{align}
\f{\rp{m_{t+1} - m_t}^2 - \rp{m_{t}-m_{t-1}}^2}{2} = \f{1}{2}(m_{t+1}-m_{t-1})k(m_t)
\end{align}
We now sum between two times $t_1$ and $t_2$:
\begin{align}
\f{1}{2}\rp{\sp{m_{t_2+1}-m_{t_2}}^2 - \sp{m_{t_1}-m_{t_1-1}}^2} 
=& \suml{t_1}{t_2}\f{1}{2}(m_{t+1}-m_{t-1})k(m_t)
\\
=& V_{t_1}-V_{t_2}
\end{align}
with
\begin{align}
 V_t = -\suml{\tau=1}{t} \f{1}{2}(m_{\tau+1}-m_{\tau-1})k(m_\tau) \ .
\end{align}
Thus we have
\begin{align}
\label{eq:energy-conservation}
E \equiv 
\f{1}{2}\sp{m_{t_2+1}-m_{t_2}}^2 + V_{t_2} = 
 \f{1}{2}\sp{m_{t_1}-m_{t_1-1}}^2 + V_{t_1}
\end{align}
which shows conservation of energy in discrete time, $V$ playing the role of a path-dependent potential. 
Notice that if the increments $m_{\tau+1}-m_{\tau}$ are small throughout the trajectory (up to $t$), then
\begin{align}
\label{eq:potential.position}
V_t
\simeq 
-\intl{(m_{1}+m_0)/2}{(m_{t}+m_{t+1})/2}k\rp{m}\d{m}
=
V((m_t+m_{t+1})/2)-V((m_1+m_0)/2)\ .
\end{align}
Thus in the low-speed regime the dynamics are Newtonian-like with a purely position-dependent potential. Up to a constant, 
\begin{align}
\label{eq:potential.explicit}
 V(x) = x(x-f^{-1}(x)) + \b^{-1}\rp{F(f^{-1}(x)) - F(x)}
\end{align}

\begin{figure}
\centering
 \includegraphics[width=0.7\textwidth]{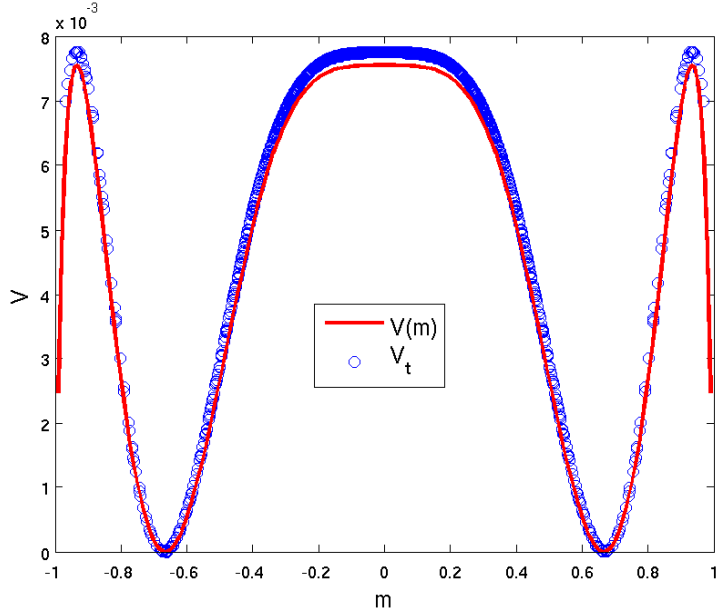}
\caption{position-dependent potential (solid line) and path-dependent potential (circles) with $T=20$ for the magnetic regime at $\b=2.5, h=0.4$ and $r_0 = 0$}
\label{fig:energy}
\end{figure}

We plot the position-dependent potential and the path-dependent potential in fig.\ref{fig:energy}. 
We obtain the trajectory by solving eqs.~(\ref{eq.motion.1}-\ref{eq.motion.2}) with $m_T$ fixed in a range of values in $[-1,1]$ using a numerical nonlinear equation solver. 
The path-dependent potential for a trajectory up to $m_{t}$ is plotted as a function of $m = (m_{t}+m_{t+1})/2$, following the convention used in \eqr{eq:potential.position}. 
We see that the position-dependent potential is a good approximation for the path-dependent potential.

\subsection{Influence of initial conditions}
\label{sec:initial}
Since the dynamics is second-order, two boundary conditions are required to specify a solution. 
Notice that \eqr{eq.motion.1}, which we can rewrite as
\begin{equation}
 m_1-m_0 = \t{f}_0(m_0)-m_0 \ ,
\end{equation}
provides an initial condition in the form of an initial velocity field.

\begin{figure}
\centering
 \includegraphics[width=0.7\textwidth]{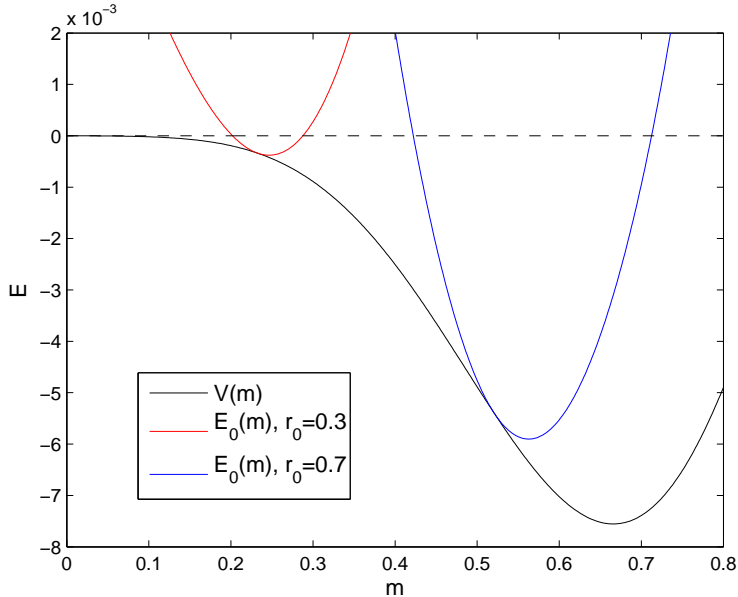}
\caption{position-dependent potential (solid black line) and initial total energy for $r_0=0.3$ (red line) and $r_0 = 0.7$ (blue) for the ferromagnetic regime at $\b=2.5$ and $h=0.4$. The level of the highest potential peak is given by the dashed line.}
\label{fig:initial_energy}
\end{figure}

We can therefore compute and plot the initial total energy of a trajectory as a function of $m_0$ in the continuum approximation of \eqr{eq:potential.position}, and compare it with the position-dependent potential. We do so in fig.~\ref{fig:initial_energy},

If $E>V_{c}$, where $V_c$ is the level of the rightmost potential energy peak, the trajectory is unbounded. But given the finite time horizon, some trajectories do not have time to leave the domain $[-1,1]$ in $T$ time steps. As $T$ increases however, the range of allowed values of $E$ above $V_{c}$ becomes narrower.
If $E<V_{c}$ on the other hand, the trajectory is bounded by the potential energy peaks. At large $T$, these trajectories will oscillate around the minimum of a potential well.

%

%


\section{Rate functions}
\subsection{Dominant trajectories}

\begin{figure}[!h]
\centering
\begin{subfigure}{0.8\textwidth}
\includegraphics[width=\textwidth]{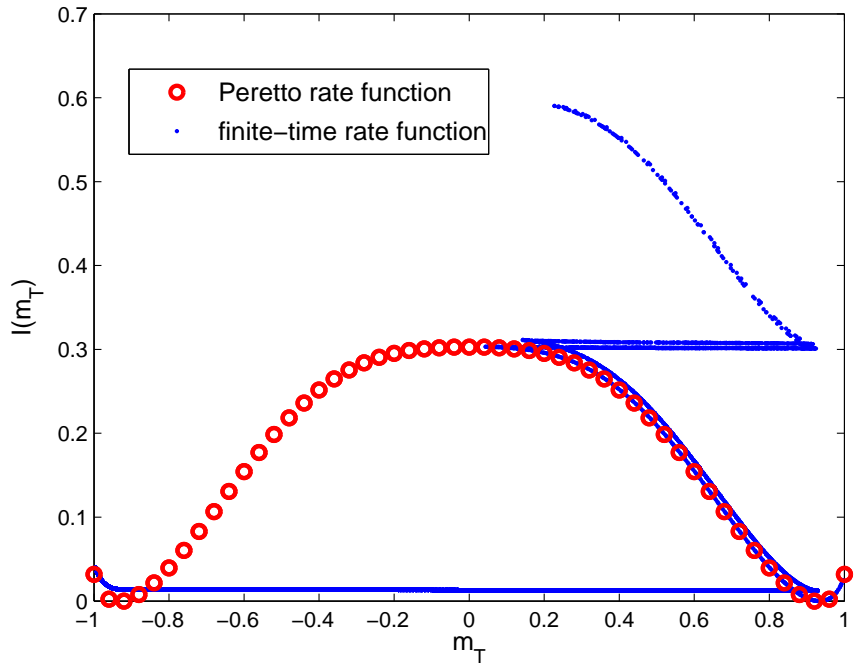}
\caption{}
\label{fig:rate.1}
\end{subfigure}
\\
\begin{subfigure}{0.8\textwidth}
\includegraphics[width = \textwidth]{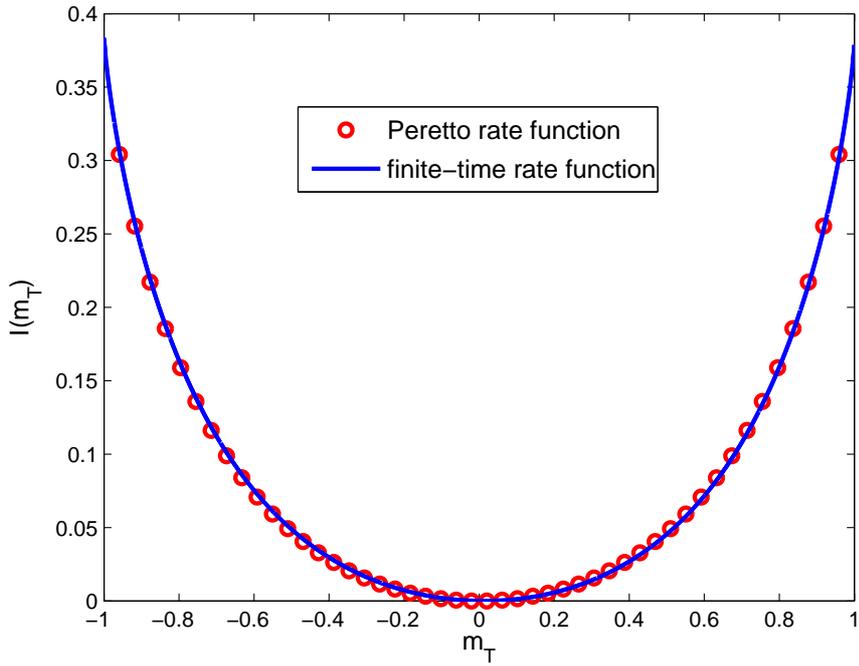}
\caption{}
\label{fig:rate.2}
\end{subfigure}
\caption{
(a) Peretto rate function (circled) and finite-time rate function ($T=50$, solid line) for the ferromagnetic phase at $\b = 2.5$, $h = 0.4, r_0 = 0.3$. 
(b) Peretto rate function (circled) and finite-time rate function ($T=20$, solid line) for the paramagnetic phase at $\b = 1$, $h = 0.4, r_0 = 0.3$. 
}
\label{fig:rate.full}
\end{figure}

\begin{figure}[!h]
\centering
\begin{subfigure}{0.8\textwidth}
\includegraphics[width=\textwidth]{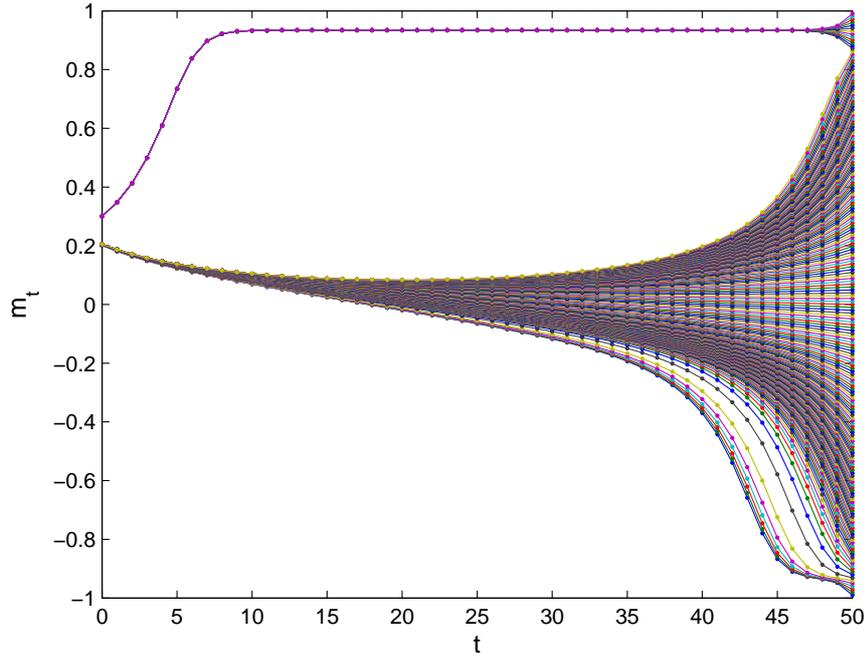}
\caption{}
\label{fig:dominant.traj}
\end{subfigure}
\\
\begin{subfigure}{0.8\textwidth}
\includegraphics[width = \textwidth]{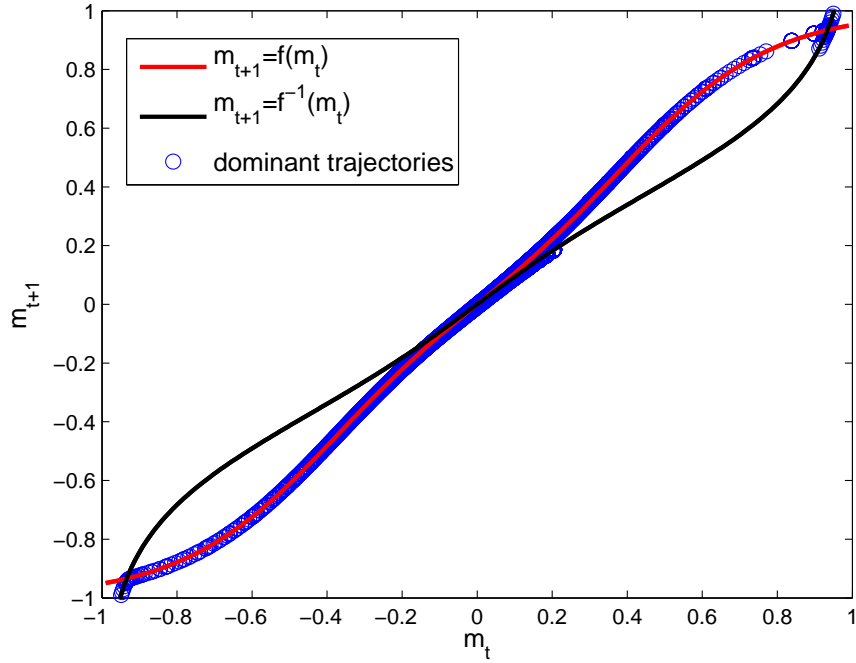}
\caption{}
\label{fig:dominant.plane}
\end{subfigure}
\caption{
(a) dominant trajectories as a function of time for $\b=2.5, h=0.4, r_0=0.3$.
\\
(b) dominant trajectories (circles) and limit curves $m_{t+1}=f(m_t)$ (solid line, red) and $m_{t+1}=f^{-1}(m_t)$ (solid line, blue) for $\b=2.5, h=0.4$.
}
\label{fig:dominant}
\end{figure}

Once we have obtained solutions to the equations of motion (\ref{eq.motion.1}-\ref{eq.motion.2}), we can compute the associated value of $\Omega$ and obtain the rate function. We plot such rate functions in fig.~\ref{fig:rate.full}.
Due to the existence of metastable trajectories, for each $m$ there are several possible trajectories ending in $m_T=m$ and thus several possible values of $\Omega$ and $I(m_T)$, as can be seen in fig.~\ref{fig:rate.1}. 
The physically relevant rate function is obtained by taking the minimum value of $I(m_T)$ at any given $m_T$. 
A natural step is then to characterize the solutions which give rise to these lowest values. We plot these trajectories in fig.~\ref{fig:dominant.traj}.

The time-dependent magnetization of the Curie-Weiss obeys the first-order dynamics $m_{t+1} = f(m_t)$. 
We know from \eqr{eq:Omega} that the contribution of this dynamics to the $\Omega$ function vanish except for initial terms, and we thus expect the dominant trajectories to follow this dynamics. 
This is well-verified in practice: as can be seen in fig.~\ref{fig:dominant.plane}, where we plot $m_{t+1}$ as a function of $m_t$ for the $\Omega$-minimizing trajectories, these trajectories verify $m_{t+1}=f(m_t)$.

There are, however, notable exceptions: at fixed $(\b,h)$, $f(x)$ takes values between $\pm f(1)$. Thus, there is no way to solve the equation $m_T = f\rp{m_{T-1}}$ if we fix $\abs{m_T}$ to be greater than $\abs{f(1)}$. 
There is however a solution to $m_T = f^{-1}(m_{T-1})$, and indeed we see the trajectories follow the time-reversed branch $m_{t+1}=f^{-1}(m_t)$ until its intersection with the forward branch. This backward dynamics comes as a cost however, hence we see the rate function $I(m)$ increase steeply past the positive fixed-point of the relaxation dynamics at $m_{eq} = 0.94$. 

\subsection{Subdominant trajectories and equilibrium}
\label{section:equilibrium}
\begin{figure}[!h]
\centering
\begin{subfigure}{0.45\textwidth}
\includegraphics[width=\textwidth]{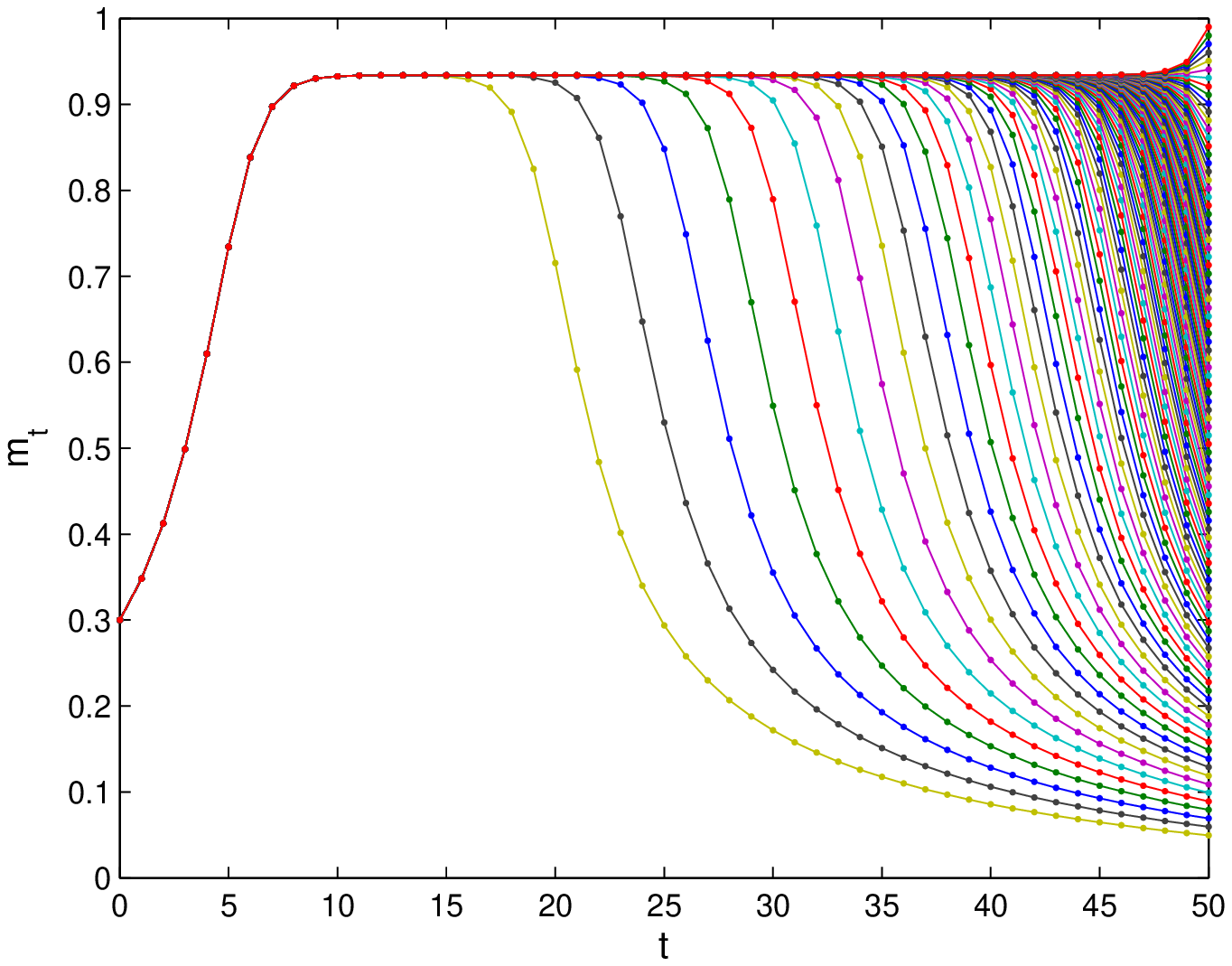}
\caption{}
\label{fig:equilibrium.traj}
\end{subfigure}
\begin{subfigure}{0.45\textwidth}
\includegraphics[width=\textwidth]{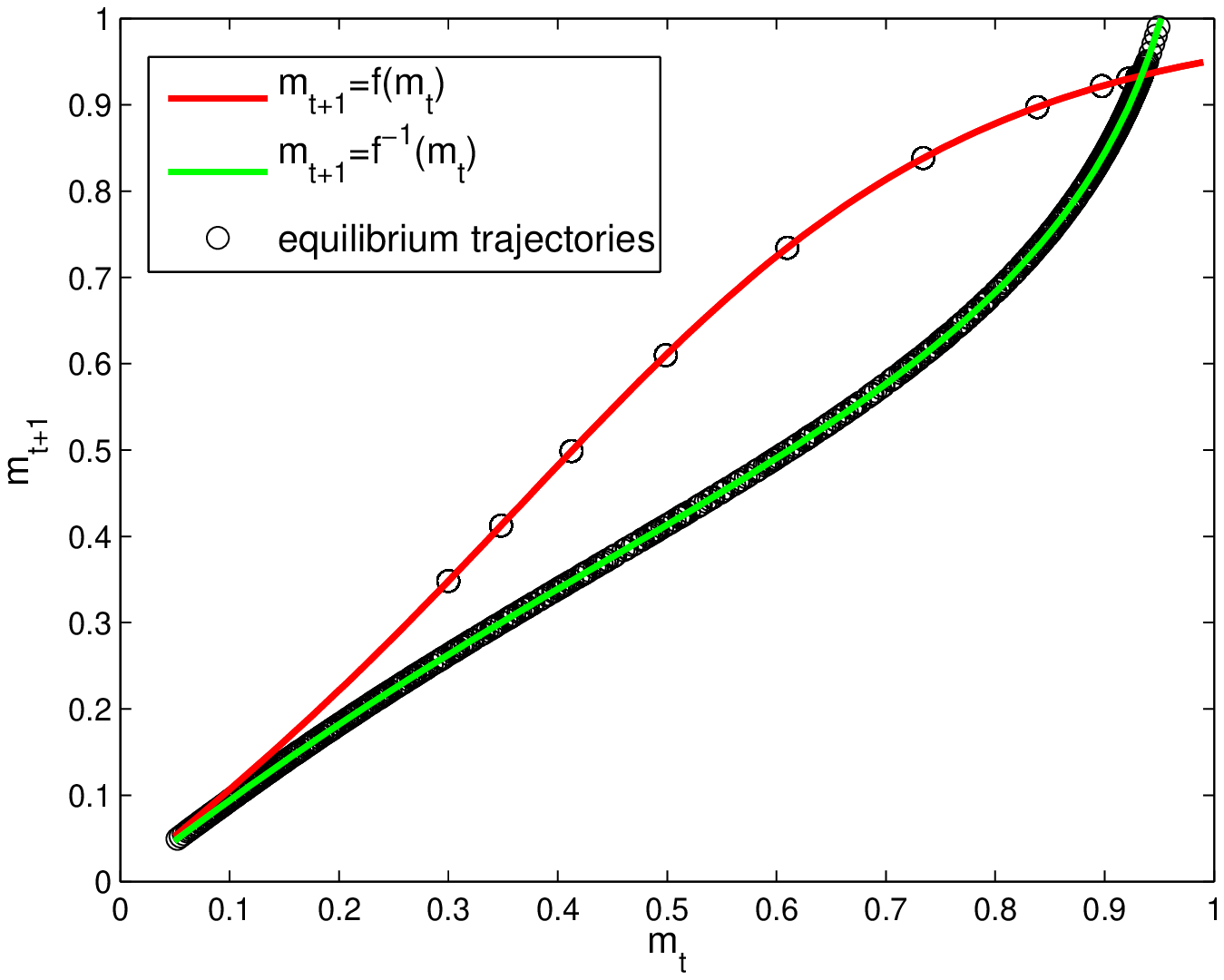}
\caption{}
\label{fig:equilibrium.plane}
\end{subfigure}
\\
\begin{subfigure}{0.45\textwidth}
\includegraphics[width=\textwidth]{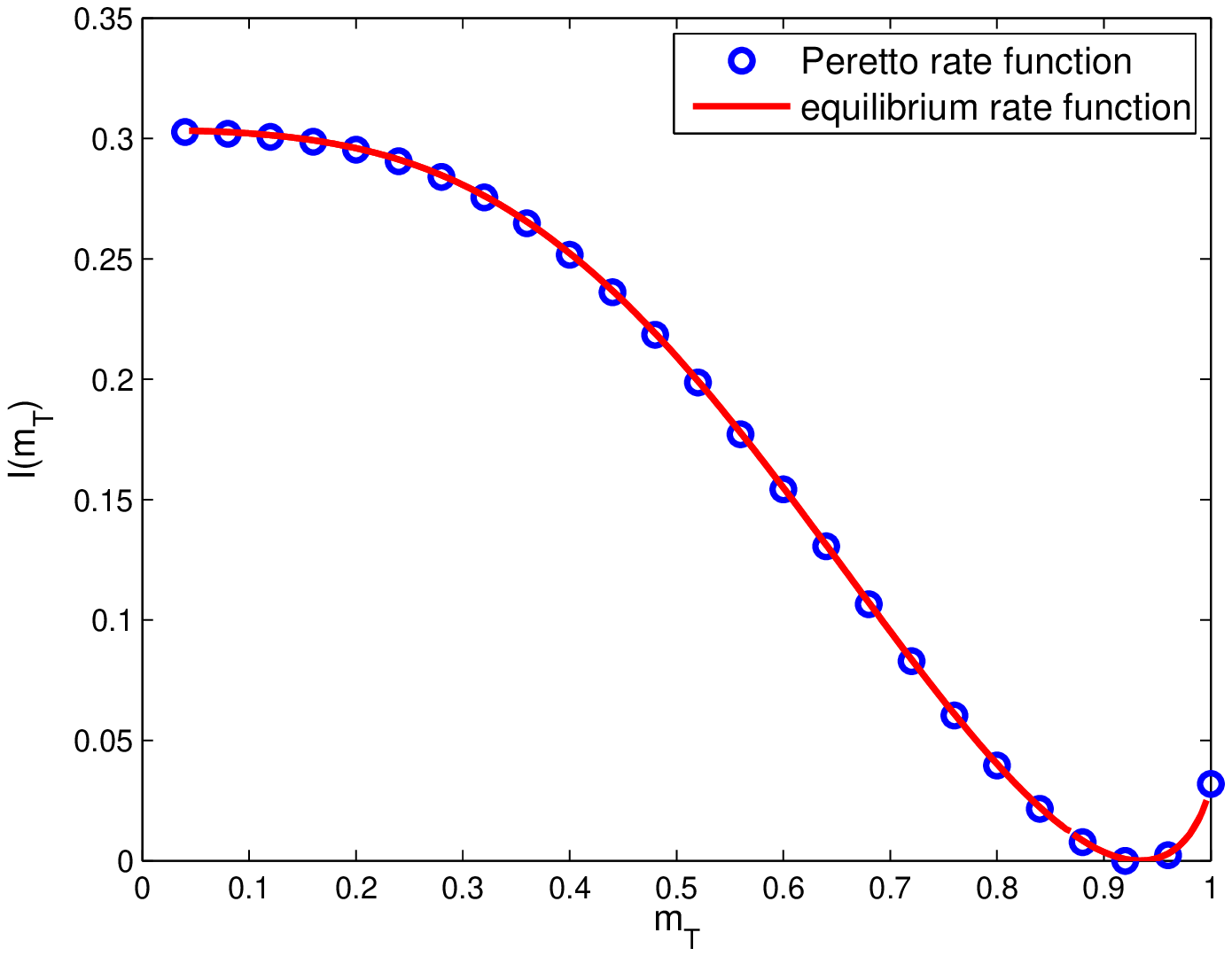}
\caption{}
\label{fig:equilibrium.rate}
\end{subfigure}
\caption{
equilibrium solutions for $\b=2.5$, $h=0.4$, $r_0=0.3$.
(a) equilibrium solutions as a function of time.
(b) quasi-solution and limit curves (circles) in the $(m_t,m_{t+1})$ plane.
(c) rate function associated with the equilibrium trajectories (red line) and Peretto rate function (circles)
}
\label{fig:equilibrium}
\end{figure}

When plotting the rate function for different types of trajectories as in \ref{fig:rate.full}, we find there is a class of trajectories whose rate function values are very close to the Peretto result \eqr{eq:Peretto}. 
We plot in fig.~\ref{fig:equilibrium.traj} these trajectories, which we call with a slight abuse of language ``equilibrium'' trajectories, as they seem to be the dominant trajectories for very large $T$. 
As can be seen in fig.~\ref{fig:equilibrium.plane}, they follow both forward ($m_{t+1}=f(m_t)$) and backwards ($m_{t+1}=f^{-1}(m_t)$) dynamics. This has a rather intuitive explanation:
we start by noting that the equation of motion (\ref{eq.motion.2})
\begin{equation}
 m_{t+1} + m_{t-1} = \t{f}\rp{m_t}
\end{equation}
is time-reversal invariant. Therefore, both forward ($m_{t+1}=f(m_t)$) and backward ($m_{t+1}=f^{-1}(m_t)$) dynamics are solution of this equation. 
But fixed-points which are stable for the forward dynamics are unstable for the backward dynamics and vice-versa.

Trajectories that follow the forward dynamics contribute minimally to the action $\Omega$. But if $T$ is large enough and if $m_T$ is not a stable fixed-point of the forward dynamics, then a dominant trajectory cannot follow forward dynamics throughout, since such a trajectory will quickly converge to a stable fixed-point $m_{eq}$ and remain there. 
A natural way for trajectories ending in $m\neq m_{eq}$ to minimise the action $\Omega$ is therefore to approach an equilibrium fixed-point via forward dynamics, and eventually veer away from it using backward dynamics to reach the prescribed value of the final magnetization.

This behaviour can be read out from fig.~\ref{fig:equilibrium.traj}, where we see several cases:
\begin{itemize}
 \item if $m_T$ is close to $m_{eq}$, the trajectories initially follow the relaxation dynamics $m_{t+1}=f(m_t)$ then veer away from $m_{eq}$ using backward dynamics.
 \item If $m_T$ is sufficiently far away from $m_{eq}$, the trajectories instead initially follow the backward dynamics $m_{t+1}=f^{-1}(m_t)$, converging toward $m=0$, then veer away from $m=0$ using forward dynamics.
 \item if $m_T<-m_{eq}$, the trajectories initially follow backward dynamics toward $m=0$ using backward dynamics, then away from $m=0$ and toward $m=-m_{eq}$ using forward dynamics, then away from $m=-m_{eq}$ using backward dynamics again.
\end{itemize}
Since this switching of dynamics happens whenever trajectories cross a stable or unstable fixed-point, more switching is possible depending on initial and terminal conditions and the number of fixed-points in the system. Up to $5$ switches can occur in the case depicted in fig.~\ref{fig:metastability} for trajectories with initial conditions to the left of the leftmost potential energy peak and terminal conditions to the right of the rightmost one.


\subsection{Stability and metastability}

\begin{figure}
\centering
 \includegraphics[width=0.7\textwidth]{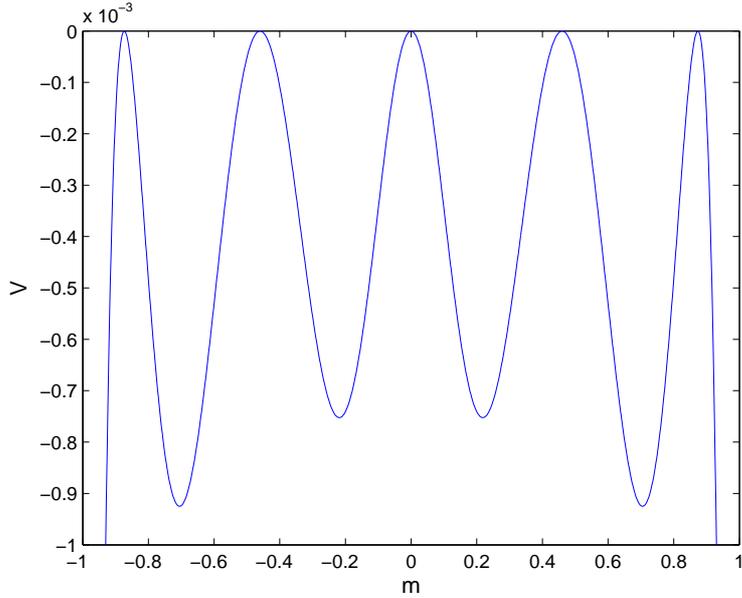}
\caption{position-dependent potential (solid black line) for $\b=2.5$ and $h=0.485$, where a ferromagnetic phase coexists with a metastable paramagnetic phase. The maxima correspond to the stable and unstable fixed points of the relaxation dynamics.}
\label{fig:metastability}
\end{figure}

In the quadruple-well potential case of $\b=2.5, h=0.485$ plotted in fig.~\ref{fig:metastability}, we can give a a good description of metastability: 
in a quadruple-well potential, if the $r_0$ parameters favors trajectories with initial magnetization $m_0$ within two peaks, one of these peaks is unstable with respect to the forward dynamics.
For example, consider a trajectory with $m_T$ at the $m=0.8$ peak with $r_0$ such that the possible initial states are between $m_0=0$ and $m_0=0.5$: the dynamics that draw closer to the peak near $m=0.5$ is necessarily backward dynamics (since it draws closer to an unstable fixed-point of the forward dynamics). 
Therefore, the uphill climb towards the peak position will incur a cost with respect to the rate function, since as was mentioned previously backward dynamics contributes more to $\Omega$ than forward dynamics. 
Thus the rate function for such a trajectory will be nonzero.
On the other hand, trajectories that draw closer to the $m=0$ stable fixed point use forward dynamics, and thus have a cost of $0$ aside from initial value contributions. 
In macroscopic terms, we would therefore see trajectories toward $m=0$ as overwhelmingly more likely than trajectories toward the positive magnetization fixed-point around $m\simeq0.8$. 
On the other hand, if the initial conditions allowed for states near the positive magnetization, the situation would be reversed and the trajectories with $m_T\simeq 0.8$ would be favored. 
Thus, a metastable equilibrium appears.

\subsection{Comparisons with simulations}

\begin{figure}[!h]
\centering
\includegraphics[width=0.8\textwidth]{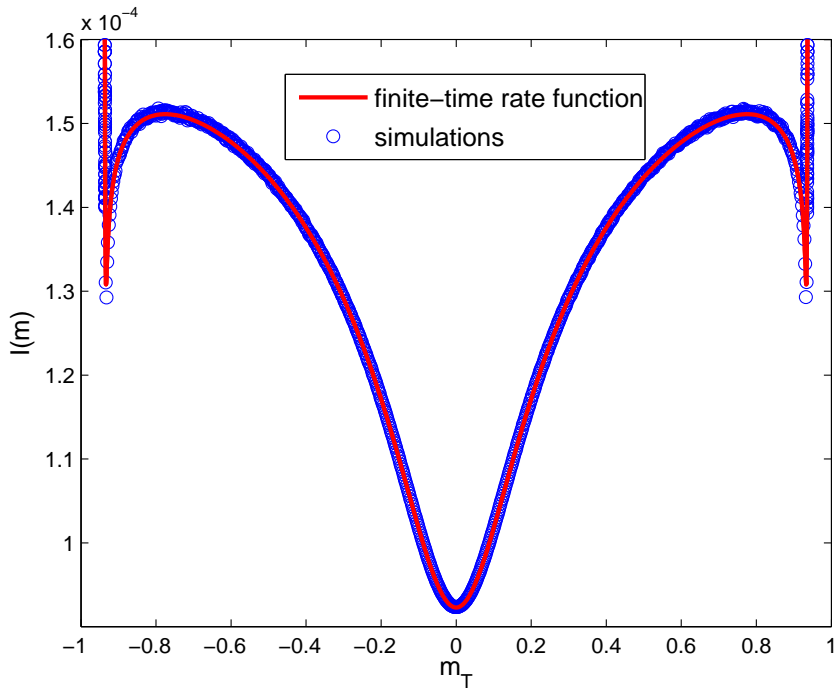}
\caption{
Finite-time ($T=50$) theoretical (solid red line) and empirical (circles) rate functions for $\b = 2.5$, $h = 0.4, r_0 = 0.0$.
}
\label{fig:temp.comparison}
\end{figure}

We run $N_{s}=10^8$ simulations with $N = 10^5$ and compare the results with analytic predictions in fig.~\ref{fig:temp.comparison}. Since we find no difference between the finite-time rate function and the Peretto result for $\b < 1$, $T=20$, we focus on the ferromagnetic part of the phase diagram and run simulations for $\b=2.5,\, h=0.4$, where the finite-time rate function and the Peretto result are highly dissimilar even as late as $T=150$. The size of the rate function is, on a large range of values of $m_T$, on the order of magnitude of $N^{-1}$. We therefore include first-order corrections to the analytically rate function as discussed in appendix \ref{first-order corrections}. We find excellent agreement. The associated Peretto rate function is plotted in fig.~\ref{fig:equilibrium.rate}, and is several orders of magnitude larger than the finite-time result.

\subsection{Connection with Langevin dynamics and Freidlin-Wentzell theory}
\label{Langevin}
The reasoning sketched in \ref{section:equilibrium} is common in Freidlin-Wentzell theory with conservative potential, where it is well-known that the trajectories realising the minimum of the action are those that follow both forward and backward dynamics (\cite{FWarticle,FWbook, BR15}).
Indeed, the above discussion is a natural adaptation of Freidlin-Wentzell theory in discrete time (a general discussion of the extension of Freidlin-Wentzell theory to Markov chains can be found in \cite{meta}): starting with \eqr{eq:proba:seq}, we see that the magnetization obtained at each time step $t$ can be decomposed into the ``deterministic'' magnetization that would be obtained in the limit of infinite system size, $m_{t}^*(m_{t-1})$ and a random error term $\delta m_{t}$, dependent on $m_{t-1}$, which if $N$ is large is Gaussian distributed via the central limit theorem, with a variance scaling like $N^{-1/2}$.

More precisely, from \eqr{eq:proba:seq} the magnetization at time $t$ is given by
\begin{align}
 m_{t} =& \f{1}{N}2\rp{b_{+}\rp{m_{t-1}} + b_{-}\rp{m_{t-1}}}-1
 \intertext{where $b_{\pm}$ are independent binomial random variables with $\tfrac{N}{2}$ trials and success probability}
 p_\pm(m_{t-1}) =& \f{e^{\b\rp{m_{t-1} \pm h}}}{2\cosh\rp{\b\rp{m_{t-1}\pm h}}}
 \intertext{In the limit of large $N$, these binomial distribution are approximately Gaussian with variance $\s(m_{t-1}) = \sqrt{2\rp{p_{+}(m_{t-1})\rp{1-p_+(m_{t-1})} + p_{-}(m_{t-1})\rp{1-p_-(m_{t-1})}}}$ and thus}
 m_{t} \simeq& f(m_{t-1}) + N^{-1/2}\s(m_{t-1}) x_t
 \intertext{where $x_t\sim \cal{N}\rp{0,1}$. We can rewrite this as}
 m_{t}-m_t \simeq& f(m_{t-1})-m_{t-1} 
 + N^{-1/2}
\s(m_{t-1}) x_t
\end{align}
i.e. the dynamics follow a discrete analogue of overdamped Langevin dynamics in an inhomogeneous medium.

\section{Conclusion}

In this article, we have studied the large deviations of the magnetization at finite times in a simple model with quenched randomness. 
The advantage of such simplicity is that a number of findings - some quite unexpected - could be studied in great analytic detail. 
Among the unexpected results, we mention in particular the emergence of a multiplicity of meta-stable trajectories, and of second-order conservative dynamics:
we find that while the dynamics of the RFIM are first-order in time, the dynamics of the relevant trajectories for large deviations are second-order and follow simple equations of motion reminiscent of Newton's second law.
In particular, they obey a form of energy conservation.
Moreover, we find that the most likely trajectories with prescribed final magnetization switch between two types of first-order dynamics: the relaxation dynamics of the Curie-Weiss RFIM, $m_{t+1} = f(m_t)$, and their time-reversed counterpart $m_{t+1}=f^{-1}(m_t)$. 
The order and time of switching is dependent on the location of temperature and field strength parameters in the RFIM phase diagram, and the initial and terminal conditions.
We study these equations numerically, obtain the relevant trajectories and compute the associated rate functions. We observe that the problem allows a multiplicity of (meta-)stable solutions, in contrast to simple relaxation dynamics.
We find very good agreement with simulations.

This mapping of the dynamics of the magnetization to that of a particle obeying some form of second-order dynamics in a multi-well potential is well-known in the study of large deviations of e.g. first passage times, as in Freidlin-Wentzell theory.
The switching of dynamics between forward and backwards has been observed in other systems as well (\cite{FWarticle, BR15, NS16}). 
To the best of our knowledge it has not been studied in the RFIM before, despite its shedding light on many surprising and counter-intuitive aspects of large deviations. 
In particular, it would be of interest to see if the switching phenomena appears in irreversible models such as studied in \cite{PK}, where the time-reversed dynamics $m_{t+1}=f^{-1}(m_t)$ cannot exist, or in the study of observables which are not local in time (e.g. average magnetization or average activity).

\section{Acknowledgements}

This work has been supported by the People Programme (Marie Curie Actions) of the European Unions Seventh Frame-
work Programme FP7/2007-2013/ under REA grant agreement n. 290038 (CF).

\pagebreak

\bibliographystyle{abbrv}
\bibliography{./RFIM_bib}{}

\appendix
\section{Equilibrium rate functions}
\label{Statics}
\subsection{RFIM Hamiltonian}
We can derive the equilibrium distribution assuming a Hamiltonian
\begin{equation}
 \cal{H}\rp{\gv{\s}} = -N\f{m^2\rp{\gv{\s}}}{2} - h\suml{i}{}\s_i \T_i \ ,
\end{equation}
as appears in \cite{lowe2013large}.

The induced distribution on the magnetization $m$ is given by
\begin{align}
 P(m) 
 =& 
 \avg{\delta\rp{m - \frac{1}{N}\suml{i}{}\s_{i}}}
 \\=&
\f{1}{Z} \suml{\cup{\s_{i}}}{}e^{\b\rp{\f{1}{2N}\rp{\suml{i}{}\s_i}^2 + \suml{i}{}\s_i\T_i}}\delta\rp{m - \frac{1}{N}\suml{i}{}\s_{i}}
 \\
 =&
\f{1}{Z} \suml{\cup{\s_{i}}}{}\exp\cup{\b\rp{N \f{m^2}{2} + \suml{i}{}\s_i\T_i}}\int \d{\hm}e^{-i\hm\rp{Nm - \suml{i}{}\s_i}}
 \\=&
\f{1}{Z} \int \d{\hm}\exp \cup{-Ni\hm m + N\b \f{m^2}{2} + N\avg{\log 2\cosh\sp{\b \T + i\hm}}_{\T}}
\end{align}
Taking the saddlepoint yields equations for real $i\hm$, which we write $\mu$, and we obtain
\begin{equation}
I^0(m) = \sup_\mu \cup{\mu m - \avg{\log \cosh\sp{\mu + \b \T }}_{\T}} - \b \f{m^2}{2}
\end{equation}
this would be the rate function, except we're missing the partition function $Z$. However $Z$ is a constant (relative to $m$), hence it can be recovered by noticing that the minimum of the rate function is $0$, hence
\begin{equation}
 I(m) = I^0(m) - \min_x\cup{I^0(x)}
\end{equation}

\subsection{Equilibrium distribution}
The previous approach, however, has the shortcoming of assuming a Hamiltonian, whereas we are more interested in obtaining the Hamiltonian from the dynamics. This has been done by Peretto (1984).
We assume parallel dynamics with one-step transition probabilities given by
\begin{align}
 W(\sv'|\sv) = \prodl{i}{}\f{\exp\cup{\s_i'\sp{\b h\T_i + \f{1}{N}\suml{j}{}\b \s_j}}}{2\cosh\sp{\b\rp{h\T_i + \f{1}{N}\suml{j}{}\s_j}}},
\end{align}
This transition probability satisfies detailed balance with an equilibrium distribution $p(\sv)$:
\begin{align}
\f{p(\sv)}{p(\sv')} =& \f{W(\sv|\sv')}{W(\sv'|\sv)}
=
\prodl{i}{}
\f{
\frac{\exp\cup{\s_i\sp{\b h\T_i + \f{1}{N}\suml{j}{}\b \s_j'}}}{2\cosh\sp{\b\rp{h\T_i + \f{1}{N}\suml{j}{}\s_j'}}}
}{
\frac{\exp\cup{\s_i'\sp{\b h\T_i + \f{1}{N}\suml{j}{}\b \s_j}}}{2\cosh\sp{\b\rp{h\T_i + \f{1}{N}\suml{j}{}\s_j}}}
}
\\
=&
\prodl{i}{}
\f{
\cosh\sp{\b\rp{h\T_i + \f{1}{N}\suml{j}{}\s_j}}\exp\cup{\s_i\b h\T_i}
}{
\cosh\sp{\b\rp{h\T_i + \f{1}{N}\suml{j}{}\s_j'}}\exp\cup{\s_i'\b h\T_i}
}
\end{align}
giving naturally
\begin{align}
 p(\sv) = \f{1}{Z}\exp\cup{\suml{i}{}\log\sp{\cosh\rp{\b\sp{h\T_i + m}}} + \s_i\b h\T_i}
\end{align}
We then work out the distribution on $m$ induced by this distribution on $\sv$ in the usual way:
\begin{align}
 P(m)
 =&
 \avg{\delta\rp{m - \frac{1}{N}\suml{i}{}\s_{i}}}
 \\=&
\f{1}{Z} \suml{\cup{\s_{i}}}{}\exp\cup{\suml{i}{}\log\sp{\cosh\rp{\b\rp{h\T_i + m}}} + \s_i\b h\T_i}\delta\rp{m - \frac{1}{N}\suml{i}{}\s_{i}}
 \\
 =&
\f{1}{Z} \suml{\cup{\s_{i}}}{}\exp\cup{\suml{i}{}\log\sp{\cosh\rp{\b\rp{h\T_i + m}}} + \s_i\b h\T_i}\int \f{\d{\hm}}{2\pi/N}e^{-i\hm\rp{Nm - \suml{i}{}\s_i}}
 \\=&
\f{1}{Z} \int \d{\hm}\f{N}{2\pi}\exp N \cup{-i\hm m + \avg{\log\sp{\cosh\rp{\b \sp{h\T + m}}}}_{\T} + \avg{\log\sp{\cosh\rp{\b h\T + i\hm}}}_{\T}}
\end{align}
and we see that the end result is
\begin{align}
I(m) = \sup_{x}\cup{xm - \avg{\log\sp{\cosh\rp{\b h\T + x}}}_{\T}} - \avg{\log\sp{\cosh\rp{\b \sp{h\T + m}}}}_{\T} + \f{1}{N}\log \sp{Z}
\end{align}
and we see again that the $\log Z$ constant only serves to ensure that $\inf_m\cup{I(m)} = 0$.

\section{Inverses}
We make extensive use of the inverses $f^{-1}(x)$ and $f_0^{-1}(x)$, and we derive them here.

The inverse of $f_0(x) = \tanh(\rho + \b x)$ presents no difficulty:
\begin{align}
 f_0^{-1}(x) = \b^{-1}\rp{\tanh^{-1}\rp{x}-\rho}
\end{align}

The inverse of $f(x) = \avg{\tanh\sp{\b\rp{x + h\T}}}_{\T}$ is a bit more complicated:
\begin{align}
 f(x) 
 =& 
 \avg{\tanh\sp{\b\rp{x + h\T}}}_{\T}
 \\=&
 \f{1}{2}\rp{\tanh\sp{\b\rp{x+h}} + \tanh\sp{\b\rp{x-h}}}
\intertext{making use of the addition formula $\tanh(x+y) = \f{\tanh(x)+\tanh(y)}{1+\tanh(x)\tanh(y)}$,}
 f(x) 
 =& 
 \f{1}{2}\rp{\f{\tanh\sp{\b x} + \tanh\sp{\b h}}{1 + \tanh\sp{\b x}\tanh\sp{\b h}} + \f{\tanh\sp{\b x} - \tanh\sp{\b h}}{1 - \tanh\sp{\b x}\tanh\sp{\b h}}}
 \intertext{we write $t_y = \tanh(\b y)$ for simplicity:}
 f(x) 
 =& 
 \f{1}{2}\rp{\f{t_x + t_h}{1 + t_xt_h} + \f{t_x - t_h}{1 - t_xt_h}}
 \\=&
  \f{1}{2}\f{\rp{t_x + t_h}\rp{1-t_xt_h} + \rp{t_x-t_h}\rp{1+t_xt_h}}{1 - t^2_xt^2_h}
  \\
  =&
  \f{1}{2}\f{\rp{t_x + t_h} - t_x^2t_h - t_xt_h^2 + t_x - t_h + t_x^2t_h - t_xt_h^2}{1 - t^2_xt^2_h}
  \\
  =&
 \f{t_x\rp{1-t_h^2}}{1 - t^2_xt^2_h}
\end{align}
Thus
\begin{align}
t^2_x t^2_h f(x) + t_x\rp{1-t_h^2}  - f(x) = 0
\end{align}
and
\begin{align}
 t_x = -\f{\rp{1-t_h^2}}{2t_h^2f(x)} \pm \sqrt{\rp{\f{1-t_h^2}{2t_h^2 f(x)}}^2 + \f{1}{t_h^2}}
\end{align}
For the choice of branch, we consider $\b h\ll 1$:
\begin{align}
 t_x =&
 -\f{1}{2\rp{\b h}^2f(x)} \pm \sqrt{\rp{\f{1}{2\rp{\b h}^2 f(x)}}^2 + \f{1}{\rp{\b h}^2}}
 \\
 =&
\f{1}{\rp{\b h}^2}\rp{ -\f{1}{2f(x)} \pm \f{1}{2\abs{f(x)}}}
\end{align}
and since the result must be finite, we must have 
\begin{align}
 -\f{1}{2f(x)} \pm \f{1}{2\abs{f(x)}} = 0,
\end{align}
i.e. we select the $+$ branch if $f(x)>0$ and the $-$ branch if $f(x)<0$:
\begin{align}
 f^{-1}(x) =& 
 \b^{-1}\tanh^{-1}\rp{
 -\f{\rp{1-t_h^2}}{2t_h^2 x} + \mathrm{sign}(x)\sqrt{\rp{\f{1-t_h^2}{2t_h^2 x}}^2 + \f{1}{t_h^2}}
 }
 \\=&
 \b^{-1}\mathrm{sign}(x)
 \tanh^{-1}\rp{
\f{ 2\cosh\rp{\b h}^2\abs{x}}{1+\sqrt{1+(x\sinh\rp{2\b h})^2}}
 }
\end{align}

\section{First-order corrections}
\label{first-order corrections}
From the path integral formulation, we can rewrite \eqr{eq:final.magnetization.probability} as
\begin{align}
  P\rp{m_T}
 \propto&
 \int \sp{\prodl{t=0}{T-1}\d{\delta m_t}\d{\delta \hm_t}}\d{\delta \hat{m}_T} 
 \exp\cup{-N\Omega\rp{\v{m}^*+\v{\delta m},\vhm^*+\gv{\delta}\vhm,m,\hat{m}^*+\delta\hat{m}}}
\intertext{where the starred quantities represent the saddlepoint values. Expanding to second order, we have}
  P\rp{m_T}
 \propto&
 \int \sp{\prodl{t=0}{T-1}\d{\delta m_t}\d{\delta \hm_t}}\d{\delta \hat{m}_T} 
 \exp\cup{
 -N\Omega\rp{\v{m}^*,\vhm^*,m,\hat{m}^*}
 \right.\\&\left.
 -\f{N}{2}\pmat{\gv{\delta}\v{m} & \gv{\delta}\vhm}\partial^2 \Omega\rp{\v{m}^*,\vhm^*,m,\hat{m}^*} \pmat{\gv{\delta}\v{m} \\ \gv{\delta}\vhm}
 }
 \end{align}
i.e. we have a Gaussian colored noise on top of the Newtonian path $(\v{m}^*,\vhm^*)$. 
This noise scales like $N^{-1/2}$, hence taking the $N\ra\infty$ limit is equivalent to taking a zero-noise limit of the noisy (Langevin) dynamics.

This suggests an obvious correction to the numerical results: the $\partial^2\Omega$ matrix has size $2T+1$ (as $m_T$ is a parameter rather than an integration variable), hence for low values of $T$ it can be diagonalized easily. 
The first-order corrections can be computed once the relevant saddlepoint trajectories have been obtained:
\begin{align}
 \f{\delta I(m)}{N} = \f{1}{2}\suml{i=1}{2T+1}\log \abs{\lambda_i},
\end{align}
where the $\lambda_i$ are the eigenvalues of the  $\partial^2\Omega$ matrix, which for a trajectory $\v{m}$ is given by
\begin{align}
\intertext{if $t=0$}
\partial^2_{m_{0}m_{t'}} \Omega
 =&
\delta_{t',0}\b f'\rp{m_0} - \b\delta_{t',1} \ ,
\\
\partial^2_{\hat{m}_{0}\hat{m}_{t'}} \Omega
 =&
\delta_{t',0}\b^{-1}f_0'\rp{f_0^{-1}(m_0)} \ ,
\\
 \partial^2_{m_t\hat{m}_{t'}} \Omega 
=&
i\delta_{t',0}\ ,
\intertext{if $t\geq 1$}
\partial^2_{m_{t}m_{t'}} \Omega
 =&
\delta_{t,t'}\b f'\rp{m_t} - \b\rp{\delta_{t',t+1} + \delta_{t',t-1}} \ ,
\\
 \partial^2_{\hat{m}_{t}\hat{m}_{t'}} \Omega
 =&
\delta_{t,t'}\b^{-1}f'\rp{f^{-1}(m_t)} \ ,
\\
 \partial^2_{m_t\hat{m}_{t'}} \Omega 
=&
i\delta_{t,t'}\ .
\end{align}
We note that any constant (in $m$) factors can be obtained simply by computing the probability distribution $P(m) = \exp\cup{-N\rp{I(m) + \delta I}}$ and requiring normalisation.

\end{document}

%% file: header.tex
\usepackage{amsmath} 
\usepackage{amsthm} 
\usepackage{amssymb}	
\usepackage{setspace}
\usepackage{mathtools}
\usepackage{slashed}
\usepackage{stmaryrd}
\makeatletter 
\renewcommand{\maketitle} 
{ \begingroup \vskip 10pt \begin{center} \large {\bf \@title}
	\vskip 10pt \large \@author \hskip 20pt \@date \end{center}
  \vskip 10pt \endgroup \setcounter{footnote}{0} }
\makeatother 

\renewcommand{\v}[1]{\ensuremath{\mathbf{#1}}} 
\newcommand{\suml}[2]{\sum \limits_{\substack{#1}}^{#2}}
\newcommand{\prodl}[2]{\prod \limits_{\substack{#1}}^{#2}}
\newcommand{\gv}[1]{\ensuremath{\mbox{\boldmath$ #1 $}}} 
\newcommand{\abs}[1]{\left| #1 \right|} 
\newcommand{\avg}[1]{\left< #1 \right>} 

\renewcommand{\d}[1]{\mathrm{d}#1\,} 
\renewcommand{\sp}[1]{\left[#1\right]}
\renewcommand{\cup}[1]{\left\{#1\right\}}
\renewcommand{\cal}[1]{\mathcal{#1}}

\renewcommand{\t}[1]{\tilde{#1}}

\newcommand{\rp}[1]{\left( #1 \right)}
\newcommand{\f}[2]{\dfrac{#1}{#2}}

\newcommand{\ra}{\rightarrow}

\newcommand{\intl}[2]{\int_{#1}^{#2}}

\newcommand{\uos}[3]{\underset{#1}{\overset{#2}{#3}}}
\let\baraccent=\= 
\renewcommand{\=}[1]{\stackrel{#1}{=}} 

\theoremstyle{definition}

\theoremstyle{remark}

\newcommand{\pmat}[1]{\begin{pmatrix} #1 \end{pmatrix}}

\renewcommand{\b}{\ensuremath{\beta}}

\newcommand{\s}{\ensuremath{\sigma}}

\def\XXint#1#2#3{{\setbox0=\hbox{$#1{#2#3}{\int}$}
     \vcenter{\hbox{$#2#3$}}\kern-.5\wd0}}
